\documentclass[twocolumn,tighten]{aastex631}
\revised{\today}
\submitjournal{ApJ}
\usepackage{appendix} 
\graphicspath{{./}{figures/}}

\newcommand{\itbf}[1]{{\textit{\textbf{#1}}}}

\shorttitle{Parameters of AGN light curves}
\shortauthors{Hu et al.}

\begin{document}

\title{How can the optical variation properties of active galactic nuclei be unbiasedly measured?}

\email{tie@mail.ustc.edu.cn; zcai@ustc.edu.cn}

\author[0009-0006-9884-6128]{Xu-Fan Hu}
\affiliation{Department of Astronomy, University of Science and Technology of China, Hefei 230026, China}
\affiliation{School of Astronomy and Space Science, University of Science and Technology of China, Hefei 230026, China}

\author[0000-0002-4223-2198]{Zhen-Yi Cai}
\affiliation{Department of Astronomy, University of Science and Technology of China, Hefei 230026, China}
\affiliation{School of Astronomy and Space Science, University of Science and Technology of China, Hefei 230026, China}

\author[0000-0002-4419-6434]{Jun-Xian Wang}
\affiliation{Department of Astronomy, University of Science and Technology of China, Hefei 230026, China}
\affiliation{School of Astronomy and Space Science, University of Science and Technology of China, Hefei 230026, China}

\begin{abstract}

The variability of active galactic nuclei (AGNs) is ubiquitous but has not yet been understood. Measuring the optical variation properties of AGNs, such as variation timescale and amplitude, and then correlating them with their fundamental physical parameters, have long served as a critical way of exploring the origin of AGN variability and the associated physics of the accretion process in AGNs. Obtaining accurate variation properties of AGNs is thus essential. It has been found that the damped random walk (DRW) process can well describe the AGN optical variation, however, there is a controversy over how long a minimal monitoring baseline is required to obtain unbiased variation properties. In this work, we settle the controversy by exhaustively scrutinizing the complex combination of assumed priors, adopted best-fit values, ensemble averaging methods, and fitting methods. Then, the newly proposed is an optimized solution where unbiased variation properties of an AGN sample possessing the same variation timescale can be obtained with a minimal baseline of about 10 times their variation timescale. Finally, the new optimized solution is used to demonstrate the positive role of time domain surveys to be conducted by the Wide Field Survey Telescope in improving constraints on AGN variation properties.

\end{abstract}


\section{Introduction} \label{sec:intro}

It is plausible that every massive galaxy contains a supermassive black hole (BH), into which gas swirls and behaves as active galactic nuclei (AGNs), including both low-luminosity AGNs and luminous quasars. Ever since the discovery of quasar, the variable nature of AGN emission has been reported. So far, its physical origin is generally explored through correlations between the variation properties, e.g., timescale ($\tau$) and amplitude ($\sigma$), of AGN light curves (LCs) and other physical parameters of AGNs, e.g., BH mass and Eddington ratio \citep{Kelly2009ApJ...698..895K,Burke2021Sci...373..789B}. However, the physical origin of AGN variability is hitherto unclear.

Regardless of the physical origin, the optical AGN LCs with timescales of months to several years are found to be well described by a first-order continuous-time auto-regressive (CAR(1)) process or more commonly Damped Random Walk (DRW) process \citep{Kelly2009ApJ...698..895K,Zu2013ApJ...765..106Z}. Note departures from the DRW process have been reported at timescales shorter than several days \citep{Mushotzky2011ApJ...743L..12M,Zu2013ApJ...765..106Z} and wavelengths of the rest-frame extreme ultraviolet \citep{Zhu2016ApJ...832...75Z}. Then, more sophisticated descriptions such as a $p$-order continuous-time auto-regressive $q$-order moving average (CARMA($p,q$)) process (\citealt{Kelly2014ApJ...788...33K} and references therein) or simply a damped harmonic oscillator (i.e., CARMA(2,1)) model (\citealt{Yu2022ApJ...936..132Y}) have been explored.
Nevertheless, parameters of the CARMA($p,q$) process, more complicated than CARMA(1,0), i.e., CAR(1) or DRW, are too difficult to be physically understood, and correlations between those parameters and physical properties of AGNs are cumbersomely interpreted. Instead, the DRW is a relatively simpler model to be interpreted than the higher-order CARMA descriptions, and the two major parameters ($\tau$ and $\sigma$) of the DRW process could be, though controversial, attributed to specific physical meanings: $\tau$, also coined by the damping timescale, may indicate the typical time required for the AGN LC to become roughly uncorrelated, while $\sigma^2$ the long-term (i.e., time interval $\Delta t \gg \tau$) variance of the AGN LC \citep{Kelly2009ApJ...698..895K}. 
Therefore, the DRW process is still widely assumed to fit or simulate the AGN LCs \citep{MacLeod2010ApJ...721.1014M,Kozlowski2016MNRAS.459.2787K,Kozlowski2017,Kozlowski2021AcA....71..103K,Hu2020AJ....160..265H,Suberlak2021ApJ...907...96S,Kovačević2021MNRAS.505.5012K,Stone2022MNRAS.514..164S}, to select AGN candidates via variability \citep{Kozlowski2010ApJ...708..927K,Lei2022RAA....22b5004L} as well as to simulate the thermal fluctuations of accretion disk \citep{Dexter2011ApJ...727L..24D,Cai2016ApJ...826....7C,Cai2018ApJ...855..117C,Cai2020ApJ...892...63C}.

Thus, accurately measuring the variation properties of AGN LCs is essential, but many observational conditions, such as limited baseline and sparse or irregular sampling, can significantly affect measuring the variation properties of AGNs, especially $\tau$. Therefore, DRW simulations have been extensively performed to assess to what extent we can confidently retrieve the input intrinsic variation properties of AGNs. However, confusing conclusions have been made. For example, to obtain without systematic bias an output measured timescale, $\tau_{\rm out}$, of an observed AGN LC, \citet{Kozlowski2017} suggest the observed baseline must be at least 10 times longer than the intrinsic timescale, $\tau_{\rm in}$, but \citet{Suberlak2021ApJ...907...96S} claim a factor of  $\sim 3 - 5$ is already sufficient. Oppositely, \citet{Kozlowski2021AcA....71..103K} demonstrate a baseline of more than $30 \tau_{\rm in}$ is indispensable.

To end with these conclusions, except for the same assumption on the DRW process, there are more distinct statistical assumptions, including priors on the DRW parameters and estimators as the ``best-fit'' values for the DRW parameters. Therefore, to exhaustively explore the effects of all these selections and the origin of the aforementioned confusing conclusions, we in this work revisit the baseline required to accurately retrieve $\tau_{\rm in}$ by comparing different priors and estimators in Section~\ref{sec:style}.
Effects of photometric uncertainty, sampling cadence, and season gap on retrieving the DRW parameters are considered in Section~\ref{sect:more_obs_factors}, before highlighting the role of the 2.5-meter Wide Field Survey Telescope \citep[WFST;][]{WFST2023} in confining the variation properties of AGNs.
Finally, a brief summary is presented in Section~\ref{sect:summary}. 
The data and codes used in this work are available on Harvard Dataverse: \dataset[doi:10.7910/DVN/I4XWGU]{https://doi.org/10.7910/DVN/I4XWGU}.

\section{DRW Simulation and Fitting} \label{sec:style}

\subsection{Simulating AGN LCs as a DRW process}

Suppose $X(t)$ is an AGN LC in magnitude as a function of epoch $t$, which can be described by the DRW process following the It$\hat{\rm o}$ differential equation \citep{Kelly2009ApJ...698..895K,brockwell2016} as
\begin{equation}\label{eq:drw_diff}
    dX(t) = - \frac{1}{\tau} X(t) dt + \sigma_{\rm s} dB(t) + c dt,~~~\tau,~\sigma_{\rm s},~t > 0,
\end{equation}
where $\tau$ is the damping timescale, $\sigma_{\rm s}$ the characteristic amplitude of variations per day$^{1/2}$ (the short-term variances $\approx \sigma^2_{\rm s} \Delta t$ for $\Delta t \ll \tau$ and connected to the long-term one by $\sigma^2 = \tau \sigma^2_{\rm s}/2$), $B(t)$ the standard Brownian motion, and $c \tau$ the mean of $X(t)$. Practically, the value of $X(t)$ given $X(s)$ for $s < t$ is 
\begin{eqnarray}\label{eq:drw_inte}
X(t) &=& e^{-\Delta t/\tau} X(s) + c \tau (1 - e^{-\Delta t/\tau}) \nonumber \\
     &+& \sigma^2 (1 - e^{-2 \Delta t/\tau}) N(0,1)
\end{eqnarray}
where $\Delta t = t - s$ and $N(0,1)$ is the normal distribution with zero mean and one variance. We then use Equation~(\ref{eq:drw_inte}) to simulate AGN LCs. In particular, the starting point of the simulated LC is $X(0)\sim N(c\tau,\sigma^2)$. 

Note that the covariance function of the DRW process is 
\begin{equation}\label{eq:drw_cov}
    {\rm Cov}(\Delta t) = {\rm Cov}(X(t), X(s)) = \sigma^2 e^{-\Delta t / \tau}
\end{equation}
and the corresponding structure function is
\begin{equation}
    {\rm SF}(\Delta t) = \sqrt{2 \sigma^2 - 2 {\rm Cov}(\Delta t)} = {\rm SF}_\infty \sqrt{1 - e^{-\Delta t / \tau}}
\end{equation}
where the asymptotic rms variability on long timescales ${\rm SF}_\infty = \sqrt{2} \sigma$ \citep{MacLeod2010ApJ...721.1014M,Zu2013ApJ...765..106Z}.

\subsection{Simulation setting}

Globally following \citet{Kozlowski2017} and \citet{Suberlak2021ApJ...907...96S}, we fix an observed baseline of $T = 8$ yr and revisit how well the input intrinsic $\tau_{\rm in}$ relative to $T$ can be retrieved. For this purpose, we consider 61 different $\tau_{\rm in}$ evenly distributed from $0.001 T$ to $T$ in steps of $0.05$ dex. For each $\tau_{\rm in}$, we simulate $10^4$ different LCs with fixed ${\rm SF}_\infty = 0.2$ mag (or $\sigma = 0.14$ mag), and we also consider two kinds of cadences: SDSS-like and OGLE-like with $N = 60$ and $N = 445$ epochs, respectively. To mimic the real observation, simulated epochs are randomly distributed at night and, otherwise specified, only one epoch is allowed within 3 hours before and after midnight. 
For SDSS-like and OGLE-like observations, we consider fixed typical mean magnitudes, $\langle m \rangle$, for AGNs with $\langle r_{\rm SDSS} \rangle = 17$\ mag and $\langle I_{\rm OGLE} \rangle = 18$\ mag, but magnitude-dependent photometric uncertainties as \citet{Suberlak2021ApJ...907...96S}:
\begin{eqnarray}
\sigma^2_{\rm SDSS}(t) &=& 0.013^2+\exp[2(r_{\rm SDSS}(t)-23.36)], \\
\sigma^2_{\rm OGLE}(t) &=& 0.004^2+\exp[1.63(I_{\rm OGLE}(t)-22.55)].
\end{eqnarray}

Hereafter, we refer to an SDSS-observed AGN with conditions of $\langle r_{\rm SDSS} \rangle = 17$ mag, $T = 8$ yr, $N = 60$, and $\sigma_{\rm e} = \sigma_{\rm SDSS} \simeq 0.013$ mag, while an OGLE-observed one with $\langle I_{\rm OGLE} \rangle = 18$\ mag, $T = 8$ yr, $N = 445$, and $\sigma_{\rm e} = \sigma_{\rm OGLE} \simeq 0.025$ mag. Note here $\sigma_{\rm OGLE} \simeq 2 \sigma_{\rm SDSS}$.

Specifically, the simulated LC is $y(t) = s(t) + n(t)$, where $s(t)$ is the DRW process around $\langle m \rangle$ and the observational uncertainty, $n(t)$, is randomly drawn from a normal distribution, $N(0, \sigma_{\rm e}^2(t))$.

\subsection{Fitting DRW parameters}

Several libraries are available for fitting the DRW parameters, such as \textsc{javelin}\footnote{\url{https://github.com/nye17/javelin}} \citep{Zu2011ApJ...735...80Z}, \textsc{carma\_pack}\footnote{\url{https://github.com/brandonckelly/carma_pack}} \citep{Kelly2014ApJ...788...33K}, \textsc{celerite}\footnote{\url{https://pypi.org/project/celerite/}} \citep{Foreman-Mackey2017AJ....154..220F,Aigrain2023ARA&A..61..329A}, 
and \textsc{EzTao}\footnote{\url{https://pypi.org/project/eztao/}} \citep{Yu2022ascl.soft01001Y,Yu2022ApJ...936..132Y}. 
Among them, the \textsc{javelin} utilizes the deterministic \textsc{amoeba} method to maximize the so-called PRH likelihood (see Appendix~\ref{app:fit_methods}), while the others are Monte Carlo Markov Chain (MCMC) samplers over likelihood functions.
The \textsc{carma\_pack} is an MCMC sampler for inferring parameters of the CARMA process, 
the \textsc{celerite} is capable of fast modeling different kinds of one-dimensional Gaussian processes, including the DRW process, 
and the \textsc{EzTao} is an easier CARMA modeling based on \textsc{celerite}. 

In this work, we assume the DRW kernel (Equation~\ref{eq:drw_cov}) and fiducially compute the likelihood functions for the DRW parameters with \textsc{celerite}, complemented with $10^4$ MCMC samplings by \textsc{emcee}\footnote{\url{https://pypi.org/project/emcee/} \citep{Foreman-Mackey2013PASP..125..306F}}. 
We use the default ``stretch move'' algorithm in \textsc{emcee} for the actual MCMC sampling and confirm that using different MCMC sampling algorithms, such as the ``differential evolution'' or ``walk'' ones contained in \textsc{emcee}, does not alter our conclusions.

\subsection{Priors on and estimators for the DRW parameters}

\begin{figure*}[!t]
    \centering
    \includegraphics[width=0.45\textwidth]{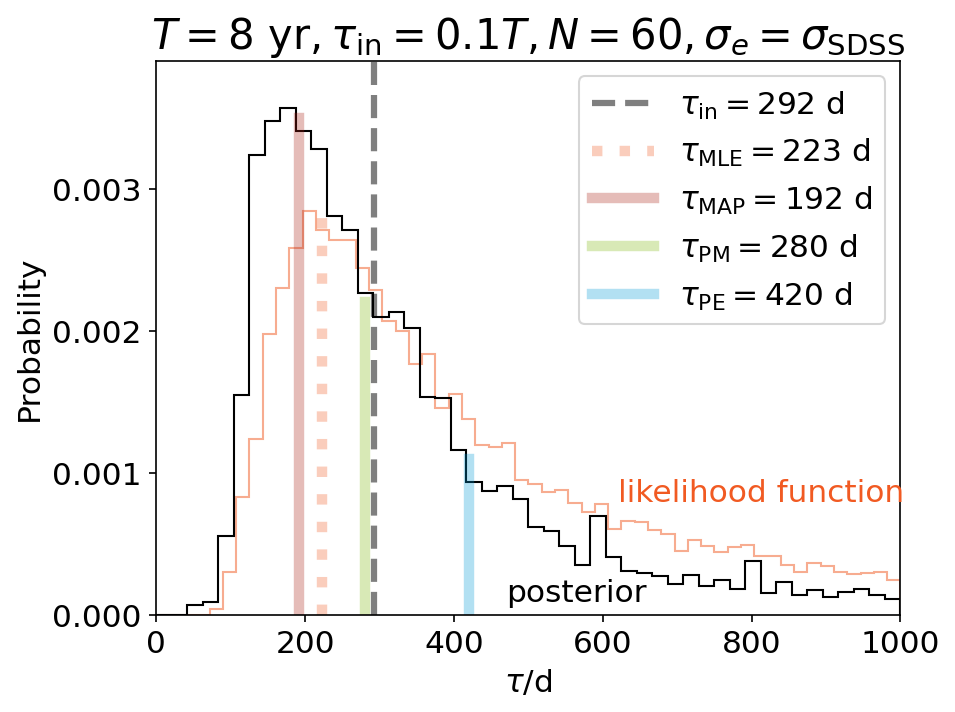}
    \includegraphics[width=0.45\textwidth]{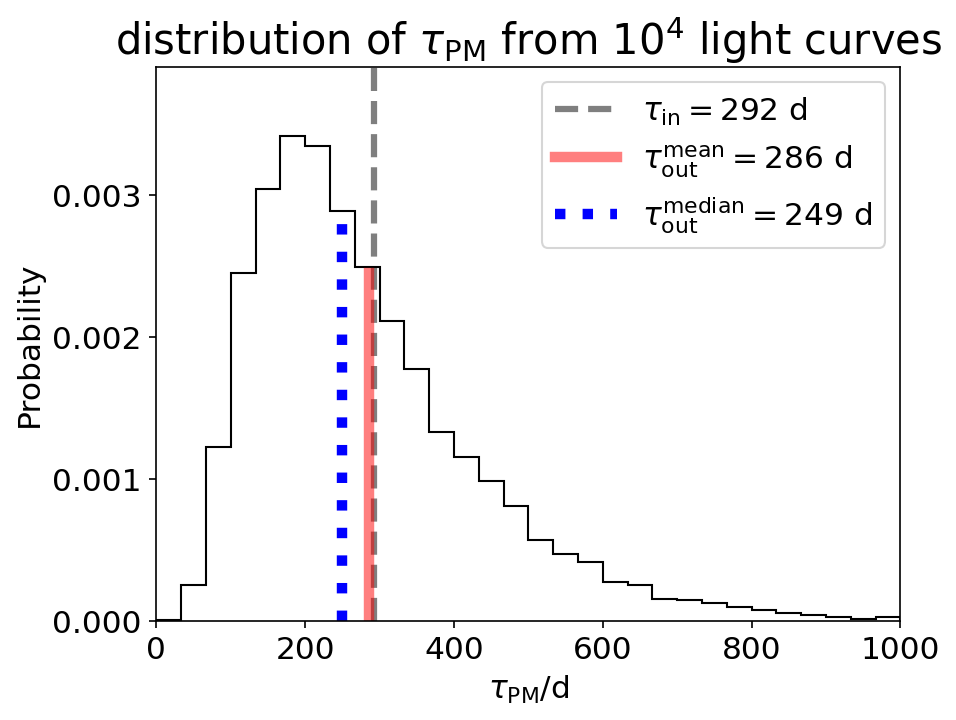}
    \caption{Left panel: given a typical LC of an SDSS-observed AGN with $\tau_{\rm in} = T / 10$ (i.e., $\tau_{\rm in} = 292$ days for given $T = 8$ years), the likelihood function for its DRW $\tau$ parameter is compared to the corresponding marginalized posterior probability distribution assuming the K17 prior, i.e., $P(\tau, \sigma) = 1/\sigma\sqrt{\tau}$. Four distinct estimators ($\tau_{\rm MLE}$ for the MLE estimator from the likelihood function, while $\tau_{\rm MAP}$, $\tau_{\rm PM}$, and $\tau_{\rm PE}$ for the MAP, PM, and PE estimators from the posterior probability distribution, respectively) are compared to the input intrinsic $\tau_{\rm in}$.
    Right panel: shown is the distribution of $\tau_{\rm PM}$ from $10^4$ LCs simulated in the same way as presented in the left panel, while the mean and median values of the distribution are compared to $\tau_{\rm in}$.}
    \label{fig:prior_estimator}
\end{figure*}

Assessing the quality of the fit and determining the ``best-fit'' value for the DRW parameters are challenging tasks. Given an observed or simulated AGN LC, we can directly estimate a potential ``best-fit'' value for its DRW parameters from the corresponding likelihood functions using the Maximum Likelihood Estimation (MLE) without priors on the DRW parameters. 
Instead, if priors are assumed, there are three potential Bayesian estimators \citep{Wei2016BayesianStatistics}, i.e., Maximum A Posterior (MAP), Posterior Expectancy (PE), and Posterior Median (PM), for the ``best-fit'' values assessed from the posterior probability distribution which is the product of the likelihood function with a prior probability distribution of the DRW parameter. Note the MAP, PE, and PM measure the peak value, the mean value (or the one-order moment), and the 50\% percentile of the posterior distribution, respectively.

Here, the MLE and MAP are obtained by maximizing the likelihood function and the posterior probability distribution, respectively, by virtue of the L-BFGS-B algorithm \citep{Byrd1995SJSC...16.1190B,Zhu1997}, while both PE and PM are derived from the posterior probability distributions constructed using $10^4$ MCMC samplings. Compared to fully explore the parameter space on fixed fine grid for the posterior probability distribution, the introduction of MCMC sampling is computational cheaper.

Previous studies have employed distinct priors and estimators for the DRW parameters and reached quite confusing conclusions. For example, \citet{Kozlowski2017} assumes priors of $P(\tau, \sigma) = {1}/{\sigma \sqrt{\tau}}$ (or equivalently $P(\tau, \sigma_{\rm s}) = {1}/{\sigma_{\rm s} \tau}$; hereafter K17 prior; see also \citealt{MacLeod2010ApJ...721.1014M,Kozlowski2010ApJ...708..927K}) and uses the MAP estimator, suggesting that a baseline longer than at least 10 times intrinsic $\tau_{\rm in}$ is required for accurately retrieving $\tau_{\rm in}$. With the same prior as \citet{Kozlowski2017}, even a baseline as long as $30\ \tau_{\rm in}$ has been suggested to be indispensable for retrieving unbiased DRW parameters \citep{Kozlowski2021AcA....71..103K}. However, \citet{Suberlak2021ApJ...907...96S} specify priors of $P(\tau, \sigma) = {1}/{\sigma \tau}$ (hereafter S21 prior) and adopt the PE estimator, claiming that a shorter baseline with only $3 - 5$ times $\tau_{\rm in}$ is sufficient for retrieving unbiased DRW parameters.

Note that both \citet{Kozlowski2017} and \citet{Suberlak2021ApJ...907...96S} take the ensemble median value of hundreds of recovered timescales from simulated LCs as the final ``best-fit'' timescale for an input $\tau_{\rm in}$.

The left panel of Figure~\ref{fig:prior_estimator} intuitively illustrates the prominent differences among aforesaid four estimators for the ``best-fit'' values of $\tau$ given a typical simulated LC of an SDSS-observed AGN with $\tau_{\rm in} = T / 10$. Therefore, for the purpose of this work, we exhaustively explore all combinations of the two priors and four estimators, as well as both the ensemble mean and median, to determine which estimator is more applicable to the assumed DRW process.

\subsection{Determining the ``best-fit'' value for the DRW parameters}

\begin{figure*}[!t]
    \centering
    \includegraphics[width=0.45\textwidth]{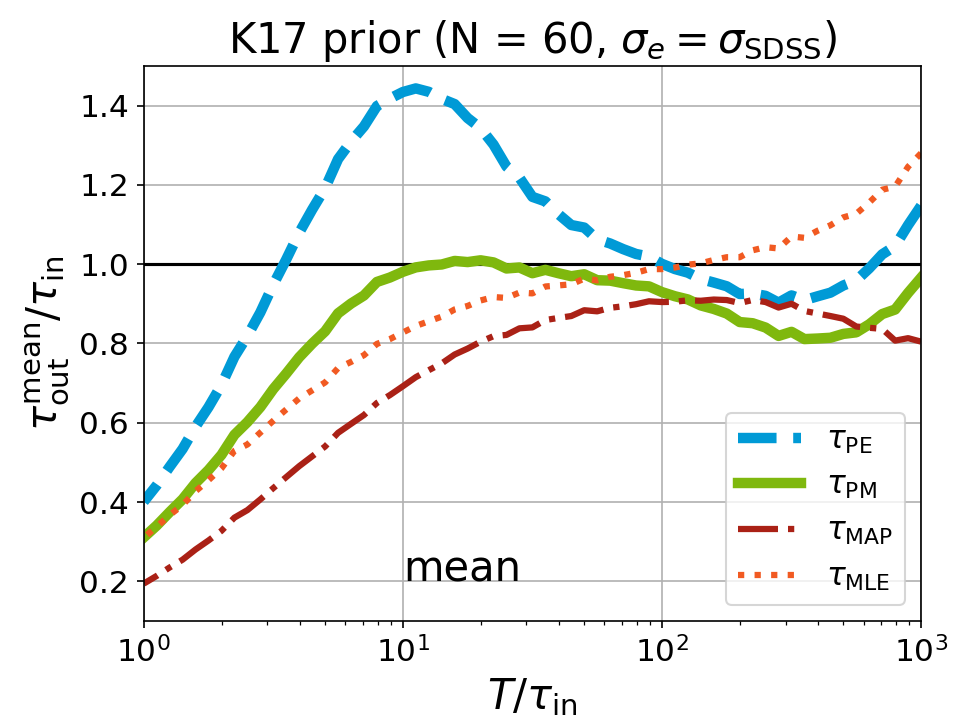}
    \includegraphics[width=0.45\textwidth]{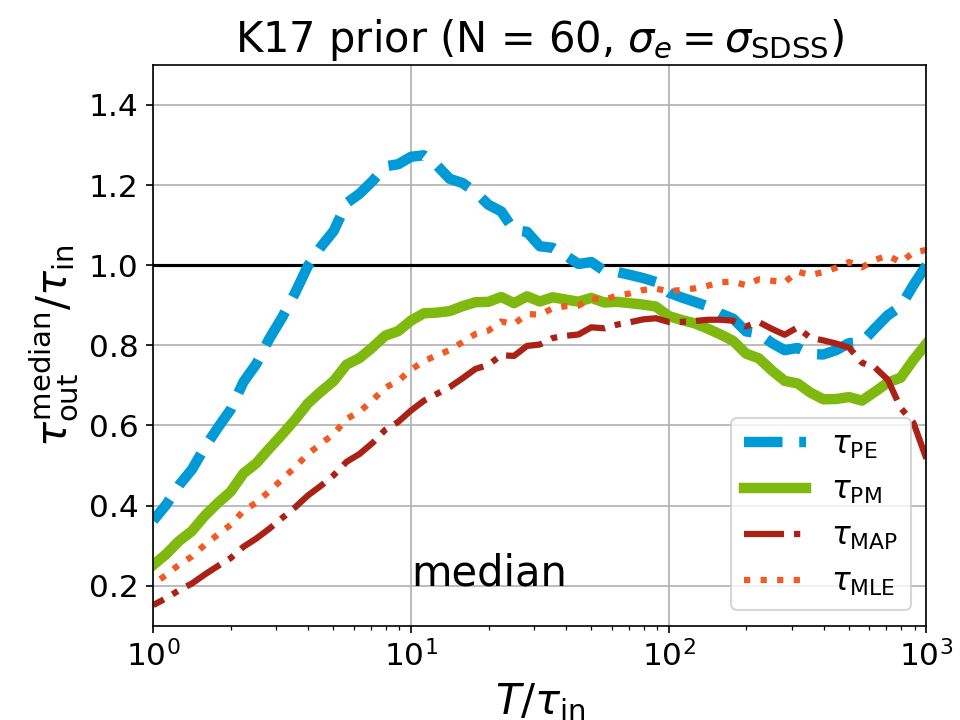}
    \includegraphics[width=0.45\textwidth]{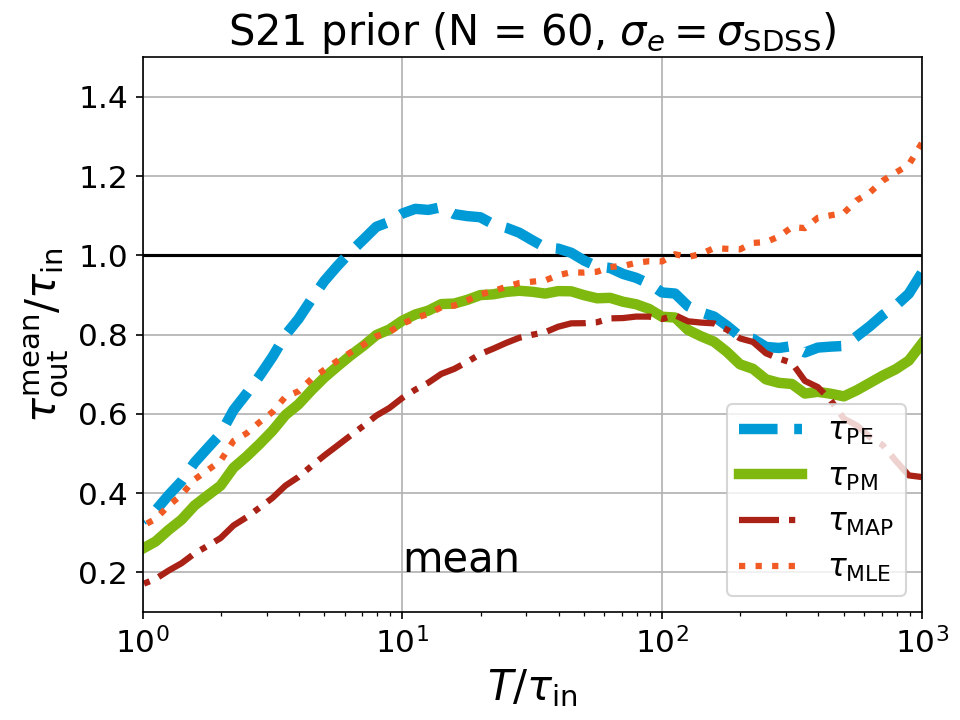}
    \includegraphics[width=0.45\textwidth]{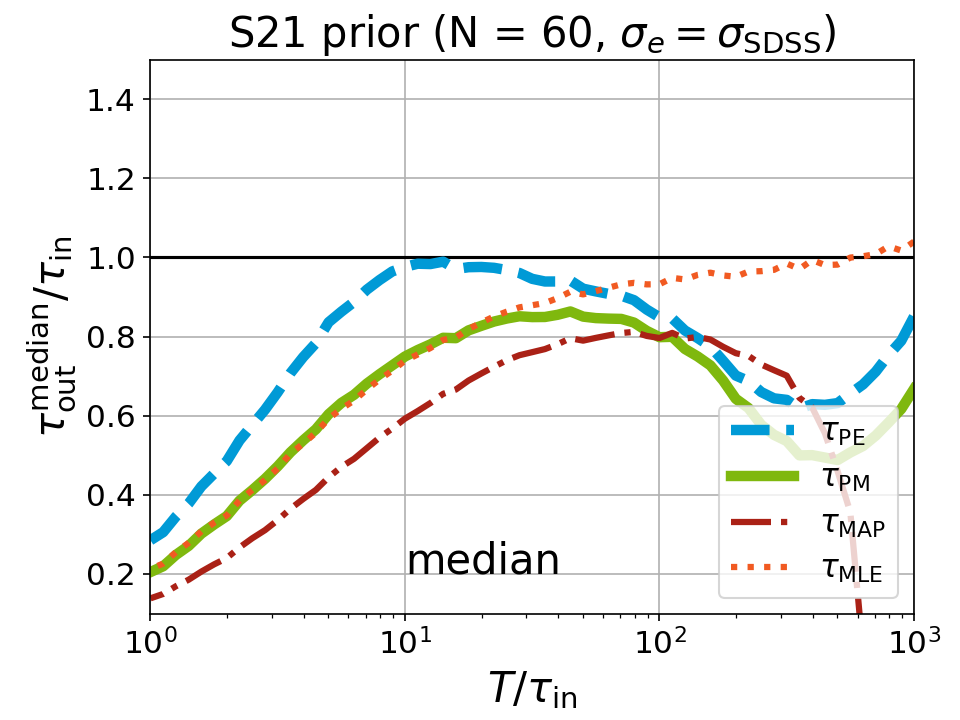}
\caption{For each given $\tau_{\rm in}$ relative to the baseline $T$ ($x$-axis), the ensemble mean/median timescales $\tau^{\rm mean/median}_{\rm out}$ relative to $\tau_{\rm in}$ ($y$-axis) for four different estimators (i.e., $\tau_{\rm PE}$, $\tau_{\rm PM}$, $\tau_{\rm MAP}$, and $\tau_{\rm MLE}$; see the left panel of Figure~\ref{fig:prior_estimator}) estimated from $10^4$ LCs of SDSS-observed AGNs are shown in the left/right panels. When estimating $\tau_{\rm PE}$, $\tau_{\rm PM}$ and $\tau_{\rm MAP}$, two priors are considered, that is, K17/S21 prior in the top/bottom panels.
}
    \label{fig:tau_out_in_n60}
\end{figure*}

\begin{figure*}[!t]
    \centering
    \includegraphics[width=0.45\textwidth]{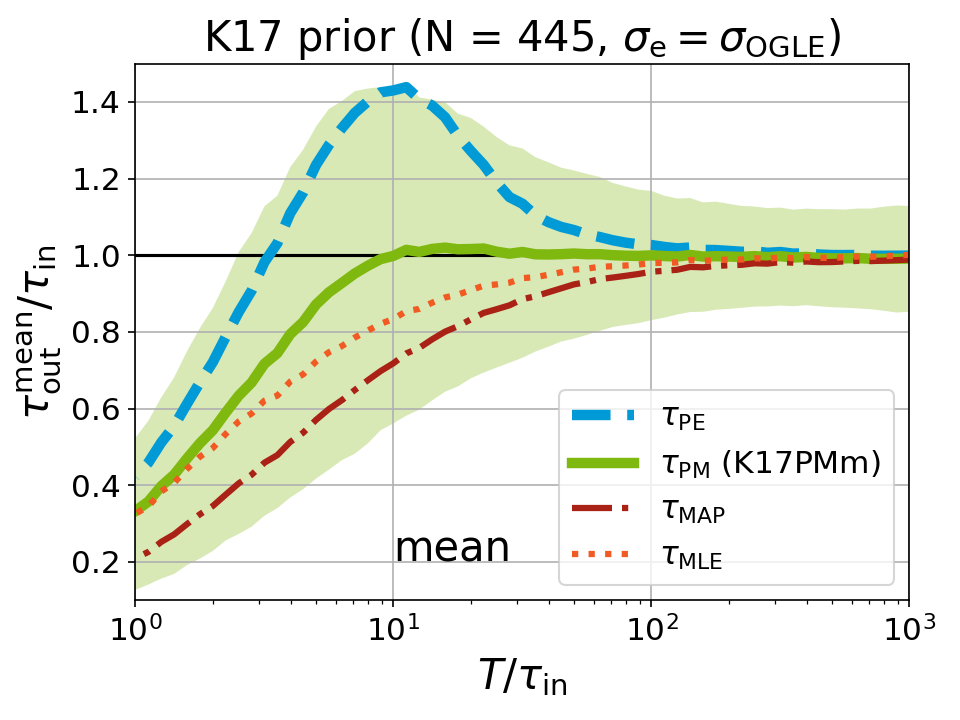}
    \includegraphics[width=0.45\textwidth]{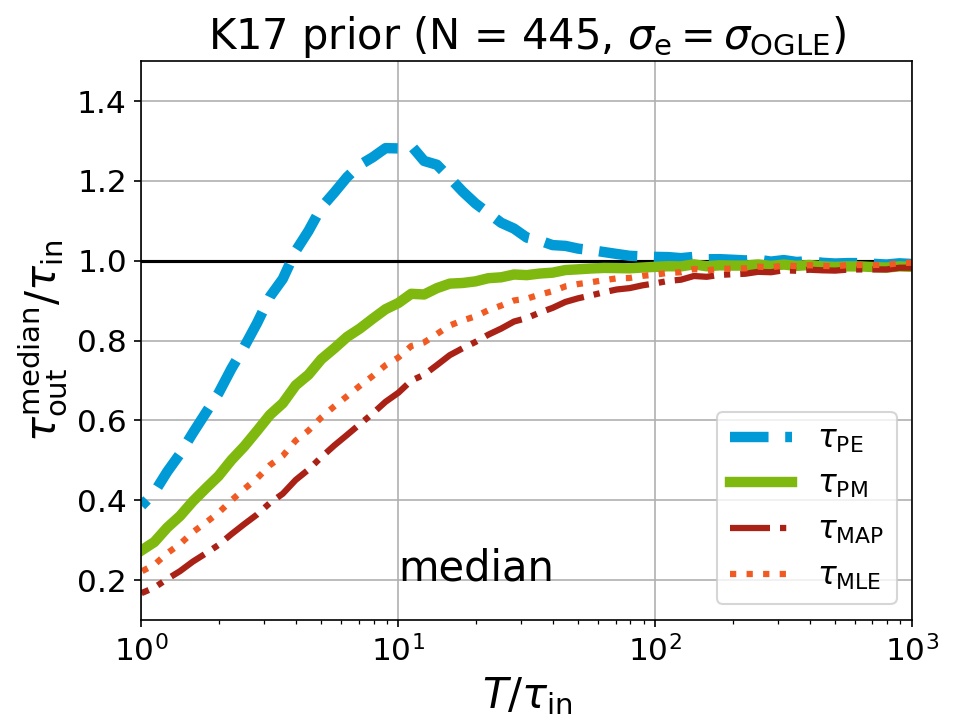}
    \includegraphics[width=0.45\textwidth]{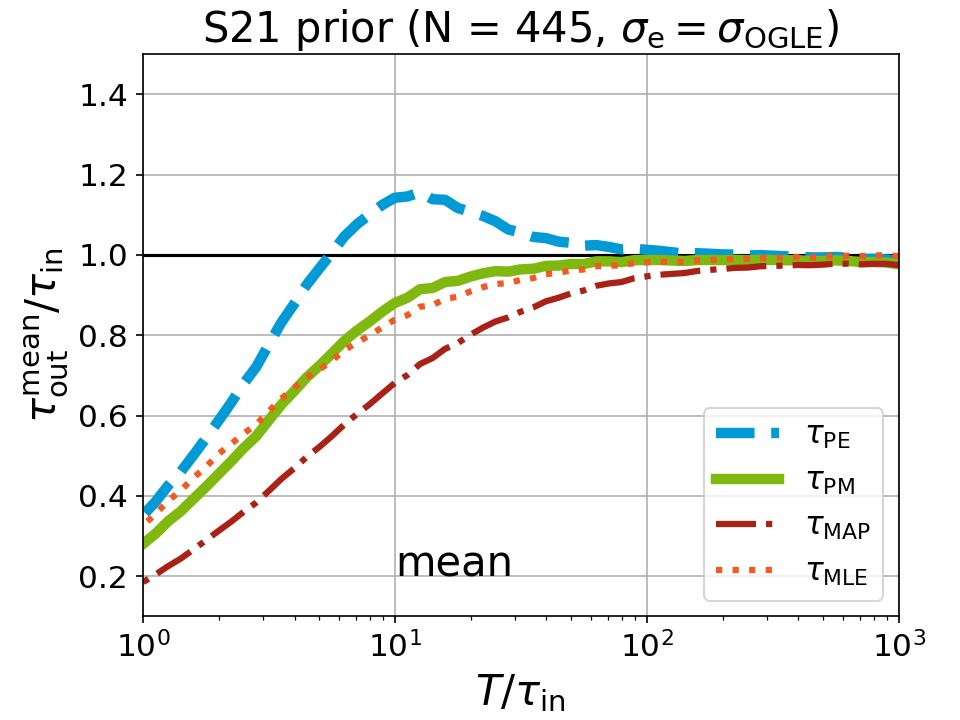}
    \includegraphics[width=0.45\textwidth]{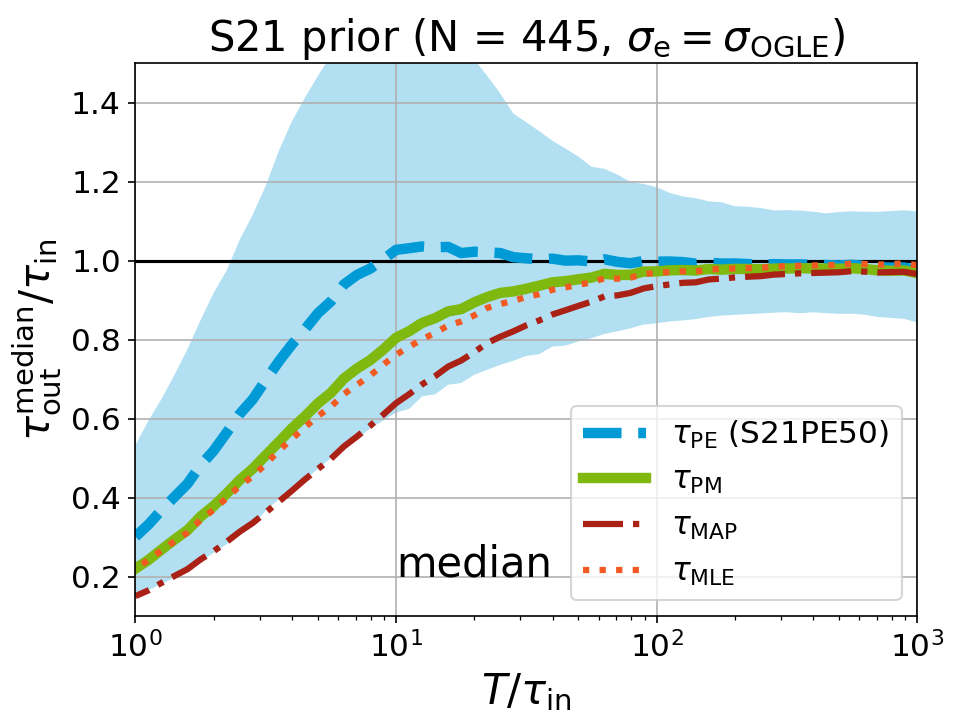}
    \caption{Same as Figure~\ref{fig:tau_out_in_n60}, but for OGLE-observed AGNs. For the two combinations that can well retrieve $\tau_{\rm in}$ down to $T / \tau_{\rm in} \simeq 10$, the 16\%-84\% percentile range of $10^4$ recovered timescales is shown as the green (blue) region for the K17 prior plus mean of $\tau_{\rm PM}$ (the S21 prior plus median of $\tau_{\rm PE}$) in the top-left (bottom-right) panel.}
    \label{fig:tau_out_in_n445}
\end{figure*}

\begin{figure}[!t]
    \centering
    \includegraphics[width=0.45\textwidth]{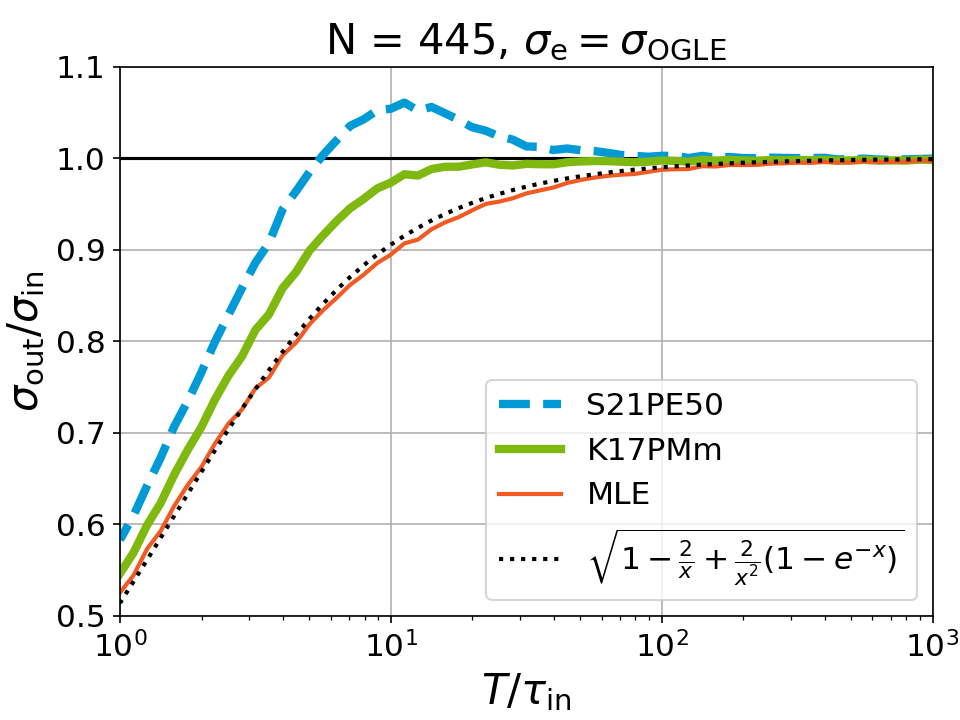}
    \caption{For OGLE-observed AGNs, $\sigma_{\rm out}$ retrieved using the K17PMm and S21PE50 solutions are compared to the mean of $\sigma_{\rm MLE}$, retrieved using MLE without prior, which is found to nicely agree with the analytical formula worked out by \citet{Kozlowski2010ApJ...708..927K} for DRW LCs, where $x = T / \tau_{\rm in}$.
    }
    \label{fig:K17_comarison_sigma}
\end{figure}

For each input $\tau_{\rm in}$, we simulate $10^4$ LCs. Given the $i$-th simulated LC, we may find several potential estimators for the ``best-fit'' value of the DRW parameters, such as $\tau^i_{\rm type}$ with ${\rm type} = $ MLE, MAP, PM, or PE for the DRW $\tau$ parameter. As illustrated in the left panel of Figure~\ref{fig:prior_estimator}, the four estimators are quite different and, for this specific LC, $\tau_{\rm PM}$ happens to be consistent with $\tau_{\rm in}$.

Owing to the randomness of AGN variability, values of the same estimator would be different from one LC to another. Therefore, we treat the ensemble mean (median) value of $\{\tau^{i=1,...,10^4}_{\rm type}\}$ as the output $\tau_{\rm out}^{\rm mean}$ ($\tau_{\rm out}^{\rm median}$) to be compared with the input $\tau_{\rm in}$. For example, the right panel of Figure~\ref{fig:prior_estimator} illustrates the probability distribution of $\tau_{\rm PM}$ retrieved from $10^4$ simulated LCs assuming the K17 prior. In this case, $\tau_{\rm out}^{\rm mean}$ is found to be in stable agreement with $\tau_{\rm in}$, while $\tau_{\rm out}^{\rm median}$ is always smaller. Note that there are individual $\tau_{\rm PM}$ much smaller (larger) than $\tau_{\rm in}$. It is found that they are typically retrieved from LCs with smaller (larger) variation amplitude, confirming the bias due to the limited baseline reported by \citet{Kozlowski2021AcA....71..103K}.

Furthermore, for a series of $\tau_{\rm in}$, Figures~\ref{fig:tau_out_in_n60} and \ref{fig:tau_out_in_n445} illustrate to what accuracy can we measure the DRW $\tau$ parameter for SDSS-observed and OGLE-observed AGNs, respectively, by considering all combinations of two priors, four estimators, and two ensemble average methods. 
Globally, estimators inferred from the S21 prior are more or less smaller than those from the K17 prior. Roughly, we find $\tau_{\rm PE} > \tau_{\rm PM} \gtrsim \tau_{\rm MLE} > \tau_{\rm MAP}$, while differences among them complicatedly depends on $T / \tau_{\rm in}$.
For the same prior and estimator, $\tau_{\rm out}^{\rm mean}$ is somewhat larger than $\tau_{\rm out}^{\rm median}$, especially around $T / \tau_{\rm in} \sim 10$. 

Comparing $\tau_{\rm out}$ to $\tau_{\rm in}$, there are prominent departures at either too long average cadence or too short baseline relative to $\tau_{\rm in}$. The effect of the former can be found in Figure~\ref{fig:tau_out_in_n60} for the SDSS-observed AGNs, whose average cadence $\Delta T$ is only $ \sim T / 60$, resulting in $\Delta T / \tau_{\rm in} > 1$ when $T / \tau_{\rm in} > 60$. 
Without priors, $\tau_{\rm MLE}/\tau_{\rm in}$ increases monotonically with increasing $T/\tau_{\rm in}$ since the information of the short-term variation is lost more significantly for larger $\Delta T$ given a fixed number of epochs. Instead, once a prior on $\tau$ is assumed, a competition between the likelihood without prior (preferring larger timescales) and the prior (preferring smaller timescales for those discussed here) results in the special dependence (like a valley\footnote{We confirm that a similar dependence does exist in terms of simulations performed using the code of \citet{Suberlak2021ApJ...907...96S}. However, the left panel of their Figure~1 does not clearly display such a dependence since their $y$-axis is $\tau_{\rm out}/T$ in logarithm rather than our $\tau_{\rm out}/\tau_{\rm in}$ in linear scale.}) of $\tau_{\rm MAP}$, $\tau_{\rm PE}$, and $\tau_{\rm PM}$ on $T/\tau_{\rm in}$ when LCs are too sparsely sampled at $T/\tau_{\rm in} > 10^2$ (Figure~\ref{fig:tau_out_in_n60}). Although by increasing the number of epochs or shortening cadence the effect of the special dependence becomes marginal (Figure~\ref{fig:tau_out_in_n445}), unavoidable is the significant departure at too short baselines with $T / \tau_{\rm in} < 10$ (Figures~\ref{fig:tau_out_in_n60} and \ref{fig:tau_out_in_n445}).
Therefore, in terms of Figure~\ref{fig:tau_out_in_n445}, we look for the best combinations that can recover $\tau_{\rm in}$ as much as possible.

As a rule of thumb, we find that there are two best combinations: K17 prior plus ensemble mean of $\tau_{\rm PM}$ (hereafter K17PMm solution) and S21 prior plus ensemble median of $\tau_{\rm PE}$ (hereafter S21PE50 solution). Both of them can quite well retrieve $\tau_{\rm in}$ down to $T / \tau_{\rm in} \simeq 10$, within uncertainties of $2\%$ and $4\%$ for the K17PMm and S21PE50 solutions, respectively (Figure~\ref{fig:tau_out_in_n445}). For other combinations, $T / \tau_{\rm in}$ as large as $\simeq 100$ is a prerequisite for achieving comparable accuracy as the two best combinations. 

Employing the MAP estimator to determine the DRW parameters, \citet{Kozlowski2021AcA....71..103K} point out that underestimated variances in shorter AGN LCs with $T / \tau_{\rm in} \lesssim 30$ lead to underestimated timescales as compared to $\tau_{\rm in}$. Contrarily, utilizing the K17PMm and S21PE50 solutions, timescales can be retrieved quiet well down to $T / \tau_{\rm in} \simeq 10$. 
Since the variance and timescale are correlated, we confirm in Figure \ref{fig:K17_comarison_sigma} that the MLE-estimated variances of our simulated LCs are nicely consistent with the prediction of DRW LCs \citep{Kozlowski2010ApJ...708..927K}, and that variances retrieved using the K17PMm and S21PE50 solutions are not significantly biased down to $T / \tau_{\rm in} \simeq 10$, especially for the K17PMm solution. Therefore, timescales down to $T / \tau_{\rm in} \simeq 10$ can be well retrieved with the K17PMm solution.

Besides less unbiased variances retrieved using the K17PMm solution down to $T / \tau_{\rm in} \simeq 10$, we also find that, for both the timescale and variance, the mean square error (MSE) of the K17PMm solution is at least half of that of the S21PE50 solution, thus the K17PMm solution is recommended as the final best way of determining the ``best-fit'' value for the DRW parameters down to $T / \tau_{\rm in} \simeq 10$.

\subsection{Comparison with previous works}

\begin{figure*}[!t]
    \centering
    \includegraphics[width=0.45\textwidth]{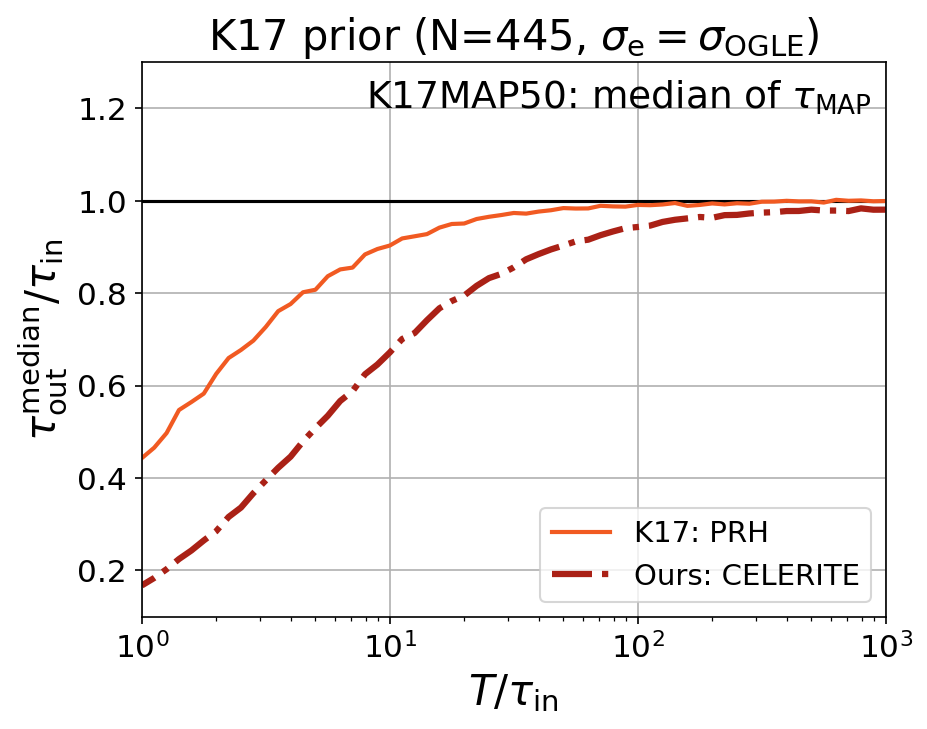}
    \includegraphics[width=0.45\textwidth]{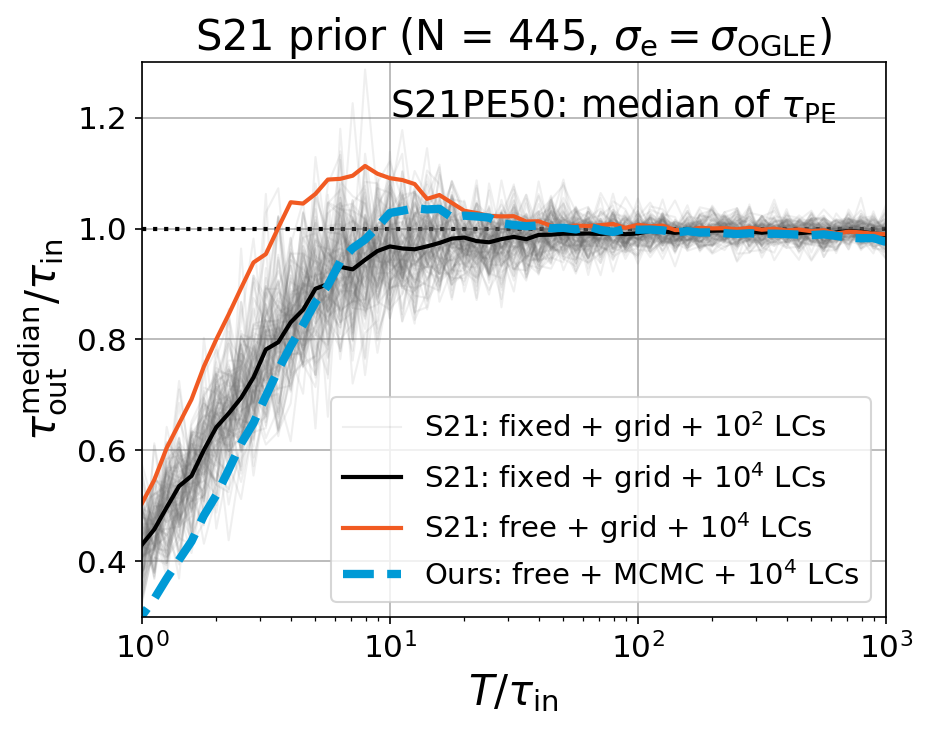}
    \caption{Left panel: considering the same simulations but different fitting methods for OGLE-observed AGNs, our ensemble median of $\tau_{\rm MAP}$ retrieved using \textsc{celerite} as a function of $T / \tau_{\rm in}$ is compared with that retrieved using the PRH method by \citet{Kozlowski2017}.
    Right panel: considering the same \textsc{celerite} but different ways of simulating LCs (i.e., fixed or free starting point to the mean of LC) and of estimating the DRW parameters (i.e., fixed grid or MCMC searching for the best parameters) for OGLE-observed AGNs, our ensemble median (based on $10^4$ LCs) of $\tau_{\rm PE}$ obtained using ``free starting point + MCMC'' is compared with that obtained using ``fixed starting point + grid'' as performed by \citet{Suberlak2021ApJ...907...96S} and that obtained using ``free starting point + grid''. For the latter two results, we have adopted the code on GitHub provided by \citet{Suberlak2021ApJ...907...96S}. Since \citet{Suberlak2021ApJ...907...96S} take the median of $\tau_{\rm PE}$ of only 100 LCs, we also illustrate 100 $\tau_{\rm out}^{\rm median}$ by taking the median of $\tau_{\rm PE}$ of 100 LCs, randomly selected from the corresponding $10^4$ LCs, simulated and fitted in the same way as \citet{Suberlak2021ApJ...907...96S}.
    }
    \label{fig:K17_S21_comparison_tau}
\end{figure*}

\subsubsection{Ours versus Koz{\l}owski (2017)}

To determine the DRW parameters, \citet{Kozlowski2017} assume the K17 prior plus ensemble median of $\tau_{\rm MAP}$ (hereafter, K17MAP50 solution) and find that a minimum of $T / \tau_{\rm in} \simeq 10$ is required for meaningfully measuring the timescale of AGN LC. However, according to the dot-dashed curve presented in the top-right panel of Figure~\ref{fig:tau_out_in_n445}, our K17MAP50 solution does not perform well at $T / \tau_{\rm in} \sim 10$, contrary to the statement of \citet{Kozlowski2017}. 
Note, rather than using \textsc{celerite} we adopted here, \citet{Kozlowski2017} utilize the so-called PRH \citep{Press1992ApJ...385..404P} method summarized in \citet{Kozlowski2010ApJ...708..927K}. This fitting method is also implemented in \textsc{javelin} (S. Koz{\l}owski 2023, personal communication; see also Appendix~\ref{app:fit_methods}).

To understand the difference between us and them, we repeat the same simulation but determine the DRW parameters using the PRH method (S. Koz{\l}owski 2023, personal communication for his PRH code).
As illustrated in the left panel of Figure~\ref{fig:K17_S21_comparison_tau}, the K17MAP50 solution involving the PRH method indeed performs much better down to $T / \tau_{\rm in} \simeq 10$, consistent with \citet{Kozlowski2017}. The slight underestimation of $\tau_{\rm out}$ at $T / \tau_{\rm in} \sim 10$ in Figure~2 of \citet{Kozlowski2017} is not so prominent owing to the fact that their $\tau_{\rm out} / T$ is presented in logarithmic scale (see also \citealt{Suberlak2021ApJ...907...96S}). Instead, presenting $\tau_{\rm out} / \tau_{\rm in}$ in linear scale as we choose here can sensitively highlight the difference between $\tau_{\rm out}$ and $\tau_{\rm in}$.
In the left panel of Figure~\ref{fig:K17_S21_comparison_tau}, from $T / \tau_{\rm in} \simeq 1$ through 10 to 100, $\tau_{\rm out}$ derived from the PRH method is larger than that from \textsc{celerite} by a factor of $\simeq 2.7$ through $\simeq 1.3$ to $\simeq 1.1$, as a result of the prominent difference in the likelihood functions and larger MAP implied by the PRH method (See Appendix~\ref{app:fit_methods}).
Therefore, we conclude that, in the sense of well retrieving the DRW parameters, the K17MAP50 solution is likely more compatible with the PRH fitting method but not \textsc{celerite}.

\subsubsection{Ours versus Suberlak et al. (2021)}

As discussed in the last subsection, the S21PE50 solution also works quite well on retrieving $\tau_{\rm in}$ down to $T / \tau_{\rm in} \simeq 10$. The same solution is adopted by \citet{Suberlak2021ApJ...907...96S}, but they suggest that $T / \tau_{\rm in} \sim 3 -5 $, shorter than what we find here, is sufficient for retrieving the timescale. Note that \citet{Suberlak2021ApJ...907...96S} state, without quantitative assessment, that as long as $T/\tau_{\rm in} \gtrsim 3$ they can recover the timescale without substantial bias. However, this is likely not true as we should demonstrate in the following.

Except adopting the same fitting method (i.e., \textsc{celerite}) and S21PE50 solution, there are still three major differences between their simulation\footnote{\url{https://github.com/suberlak/PS1_SDSS_CRTS/blob/master/code3/Fig_01_recovery_DRW_timescale.ipynb}} and ours:
\begin{enumerate}
    \item The starting points, $s(0)$, of simulated LCs are all fixed to the mean magnitude, $\langle m \rangle$, by \citet{Suberlak2021ApJ...907...96S}, while we free $s(0)$ by drawing it from a normal distribution, $N(\langle m \rangle, \sigma^2)$, mimicking a long enough ``burn-in''. 
    
    \item When estimating $\tau_{\rm PE}$, \citet{Suberlak2021ApJ...907...96S} only consider a fixed sparse grid (3600 points, that is, 60 points for $\tau$ from 1 day to 5000 days and 60 points for $\sigma$ from 0.02 to 0.7 mag, both of them are evenly spaced in logarithm) for the two DRW parameters, while we adopt the MCMC approach \citep{Foreman-Mackey2013PASP..125..306F} to approximate the posterior distributions for the two DRW parameters.
    
    \item For each $\tau_{\rm in}$, $\tau_{\rm out}$ is the median of $\tau_{\rm PE}$ retrieved from only 100 simulated LCs, while we simulate $10^4$ LCs, pursuing a stable estimation for the ensemble median.
\end{enumerate}

As illustrated in the right panel of Figure~\ref{fig:K17_S21_comparison_tau}, even both \citet{Suberlak2021ApJ...907...96S} and us have adopted the same fitting method and S21PE50 solution, differences on simulating LCs and exploring the parameter space indeed induce some prominent diversities. Using a fixed sparse grid for the DRW parameters would overestimate $\tau_{\rm out}$ as compared to using the MCMC approach, while fixing the start point would give rise to smaller retrieved $\tau_{\rm out}$. Thus, considering both a fixed grid for the DRW parameters and fixing the start point as done by \citet{Suberlak2021ApJ...907...96S} surprisingly results in $\tau_{\rm out}$ comparable to what we find, so we do not confirm their finding that $T / \tau_{\rm in} \sim 3 -5 $ is sufficient for retrieving the timescale. Instead, we show that $T / \tau_{\rm in} \simeq 10 $ or even larger is essential under the same approach as \citet{Suberlak2021ApJ...907...96S}. In the right panel of Figure~\ref{fig:K17_S21_comparison_tau}, we further demonstrate that achieving unbiased fitting with $T / \tau_{\rm in} \sim 3 -5 $ is likely attributed to the randomness of the ensemble median of $\tau_{\rm out}$, which are based on only 100 simulated LCs as done by \citet{Suberlak2021ApJ...907...96S}.

\begin{deluxetable*}{cc cc cc cc cc}
\tablenum{1}
\tablecaption{Settings of the three fitting methods and the corresponding performances \label{tab:fit_summary}}
\tablewidth{0pt}
\tablehead{
\colhead{Reference} & \colhead{Starting} & \colhead{Likelihood} & \colhead{Prior} &
\colhead{Sampling} & \colhead{Best-fit} & \colhead{Ensemble} & \colhead{Solution} & \multicolumn{2}{c}{$T/\tau_{\rm in}$} \\
\cline{9-10}
\colhead{Name} & \colhead{Point} & \colhead{} & \colhead{} & \colhead{Method} & \colhead{Estimator} & \colhead{Average} & \colhead{Abbreviation} & \colhead{2\% acc.} & \colhead{10\% acc.}
}
\decimalcolnumbers
\startdata
\citet{Kozlowski2017} & $\sim N(\langle m \rangle,\sigma^2)$ & PRH & $1/\sigma\sqrt{\tau}$ & deterministic & MAP & median & K17MAP50 & $\simeq 45$ & $\simeq 10$\\
\citet{Suberlak2021ApJ...907...96S} & $= \langle m \rangle$ & \textsc{celerite} & $1/\sigma {\tau}$ & fixed sparse grid & PE & median & S21PE50 & $\simeq 17$ & $\simeq 6$\\
Ours & $\sim N(\langle m \rangle,\sigma^2)$ & \textsc{celerite} & $1/\sigma\sqrt{\tau}$ & \textsc{emcee} & PM & mean & K17PMm & $\simeq 8$ & $\simeq 6$\\
\enddata
\tablecomments{
(1) The reference name for the fitting method proposed; 
(2) The starting point of the simulated LC: either randomly drawn from a normal distribution, $N(\langle m \rangle, \sigma^2)$, or fixed to the mean $\langle m \rangle$; 
(3) The likelihood: either PRH or \textsc{celerite} (see \hyperref[app:fit_methods]{Appendix}); 
(4) The priors for the model parameters (e.g., $\tau$ and $\sigma$): either $1/\sigma\sqrt{\tau}$ or $1/\sigma {\tau}$; 
(5) The sampling method for the model parameters: deterministic, fixed sparse grid, or MCMC sampling using \textsc{emcee}; 
(6) The estimator for the ``best-fit'' model parameter of a single LC: MAP, PE, or PM; 
(7) The method averaging the ``best-fit'' model parameters retrieved from an ensemble LCs: either median or mean;
(8) The abbreviation for the solution mentioned in this work;
(9) The minimal $T/\tau_{\rm in}$ required for achieving a 2\% accuracy for the retrieved $\tau$ and are estimated using the orange solid line in the left panel of Figure~\ref{fig:K17_S21_comparison_tau}, the black solid line in the right panel of Figure~\ref{fig:K17_S21_comparison_tau}, and the green solid line in the top-left panel of Figure~\ref{fig:tau_out_in_n445} for the \citet{Kozlowski2017}, \citet{Suberlak2021ApJ...907...96S}, and our solutions, respectively;
(10) Same as the ninth column but for achieving a 10\% accuracy.
}
\end{deluxetable*}

\section{Discussions}\label{sect:more_obs_factors}

As introduced in Section~\ref{sec:style} and summarized in Table~\ref{tab:fit_summary}, we have demonstrated that the K17PMm solution works quite well on retrieving the DRW parameters of an AGN sample down to $T / \tau_{\rm in} \simeq 10$. Here, adopting the K17PMm solution, we further consider effects of the photometric uncertainty ($\sigma_{\rm e}$), the sampling cadence (with $\langle \Delta t \rangle$ for the average cadence), and the season gap on the retrieved DRW parameters. 
Then we address how long the baseline is necessary in order to determine the DRW parameters for a single AGN down to a desired accuracy. 
Finally, time domain surveys to be conducted by the WFST \citep{WFST2023} are briefly introduced and the role of WFST surveys in improving constraints on the DRW parameters is highlighted.

In the following comparisons, we consider an AGN sample observed with a baseline of $T \gtrsim 10 \tau_{\rm in}$ and an average cadence $\langle \Delta t \rangle$. In other words, an AGN would be observed at $N$ epochs, where $N \simeq T / \langle \Delta t \rangle$. These observed epochs are randomly distributed, rather than evenly distributed, within the baseline because we confirm that for random epochs the DRW parameters can still be well retrieved even with a quite large $\langle \Delta t \rangle$ since short-term variations can be kept by random observations.
To reduce the randomness, we simulate $10^4$ LCs for each set of conditions, i.e., \{$\tau_{\rm in}$, $\sigma$, $T$, $\langle \Delta t \rangle$, $\sigma_{\rm e}$\}.

\subsection{Photometric uncertainty}

\begin{figure*}[!t]
     \centering
     \includegraphics[width=0.45\textwidth]{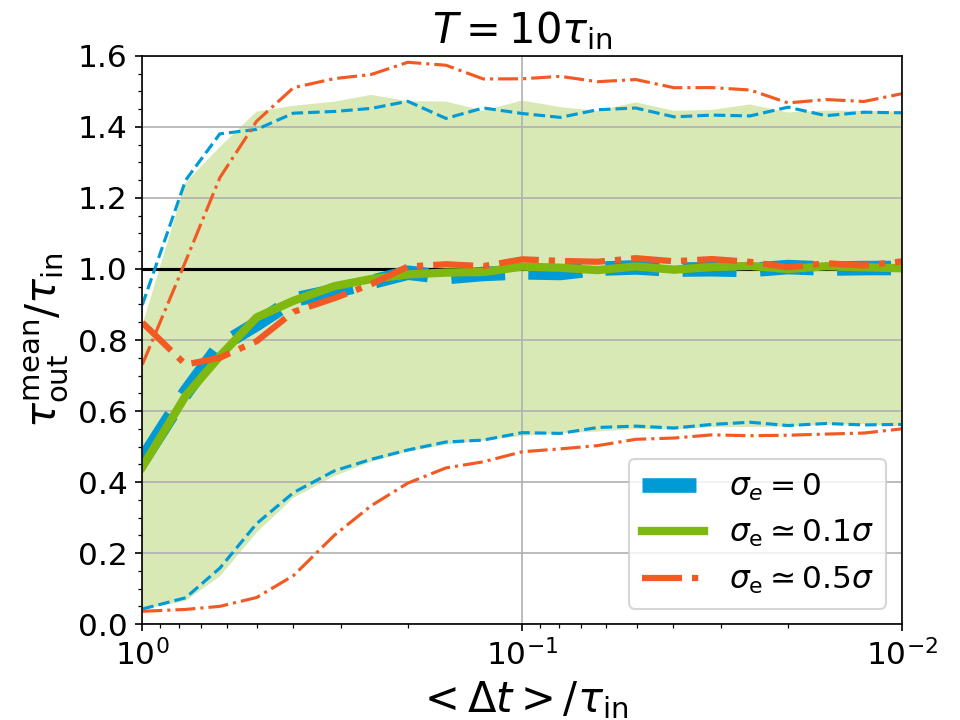}
     \includegraphics[width=0.45\textwidth]{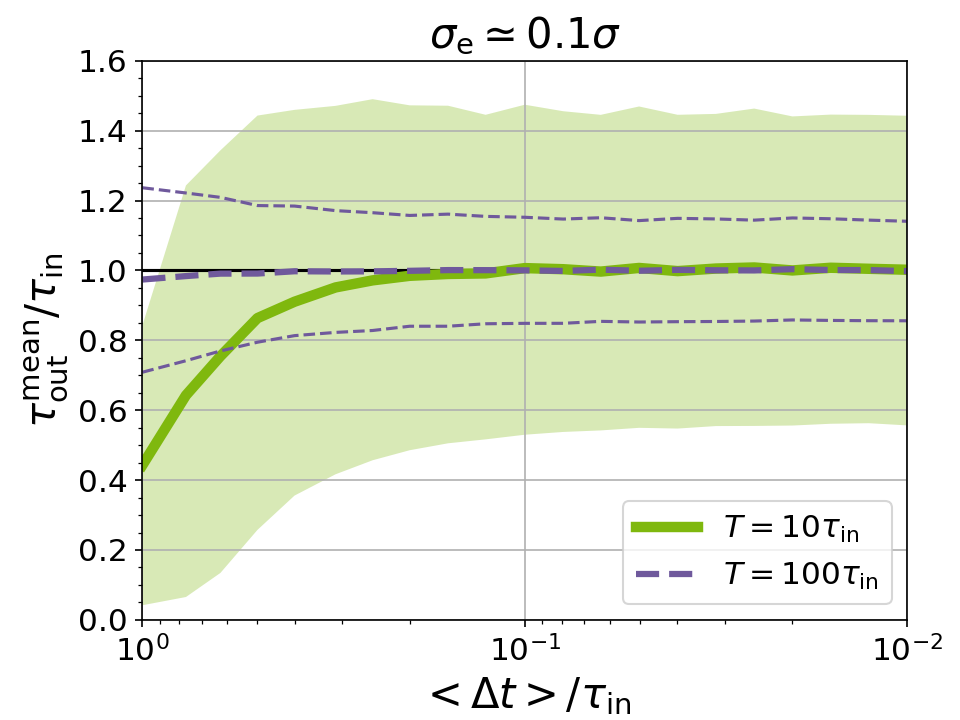}
     \caption{Left panel: adopting the K17PMm solution to retrieve the DRW timescales for an AGN sample observed with $T = 10 \tau_{\rm in}$ and different photometric uncertainties (i.e., $\sigma_{\rm e} \simeq 0$, 0.1$\sigma$, and 0.5$\sigma$, where $\sigma$ is the long-term variance of AGN LC), unbiased timescales (i.e., $\tau^{\rm mean}_{\rm out} / \tau_{\rm in} \simeq 1$) can be retrieved with the average cadence $\langle \Delta t \rangle$ shorter than $0.2 \tau_{\rm in}$, while the dispersion of the retrieved timescales (i.e., 16\%-84\% percentile range based on $10^4$ LCs) does not significantly depend on the photometric uncertainty. The 16\%-84\% percentile range for the case with $\sigma_{\rm e} \simeq 0.1 \sigma$ is indicated by the green region, while those for the other two cases are indicated by the corresponding thinner lines with the same color and line style.
     Right panel: similar to the left panel but for an AGN sample observed with $\sigma_{\rm e} \simeq 0.1 \sigma$ and $T = 10 \tau_{\rm in}$ and $100 \tau_{\rm in}$. With a longer baseline, unbiased timescales can be retrieved with larger average cadence $\langle \Delta t \rangle$ and the corresponding 16\%-84\% dispersion is narrower.
     }
     \label{fig:epochs}
 \end{figure*}

Relative to the long-term variance ($\sigma$) of AGN LC, we compare three different photometric uncertainties with $\sigma_{\rm e} = 0$, 0.1$\sigma$, and 0.5$\sigma$. For an AGN sample observed with $T = 10 \tau_{\rm in}$ and fine enough $\langle \Delta t \rangle$ (i.e., $\langle \Delta t \rangle < 0.2 \tau_{\rm in}$; see the left panel of Figure~\ref{fig:epochs}), unbiased timescales can be retrieved for all $\sigma_{\rm e}$ even though a slightly larger dispersion is found for very large $\sigma_{\rm e}$. 
Interestingly, decreasing $\sigma_{\rm e}$ from $0.1\sigma$ to 0 does not further decrease the dispersion of the retrieved timescales, suggesting that for small photometric uncertainty \textsc{celerite} can take good care of the scatter imposed by the photometric uncertainty.

In addition, the measured errors on the LCs could not be completely accurate. Thus, an extra white noise term has been introduced to account for an additional source of photometric error \citep[e.g.,][]{Burke2021Sci...373..789B,Wang2023MNRAS.521...99W}. However, we find this extra noise term has little effect on our conclusions (see Appendix~\ref{app:extra_err}).

\subsection{Sampling cadence}

\begin{figure}[!t]
     \centering
     \includegraphics[width=0.45\textwidth]{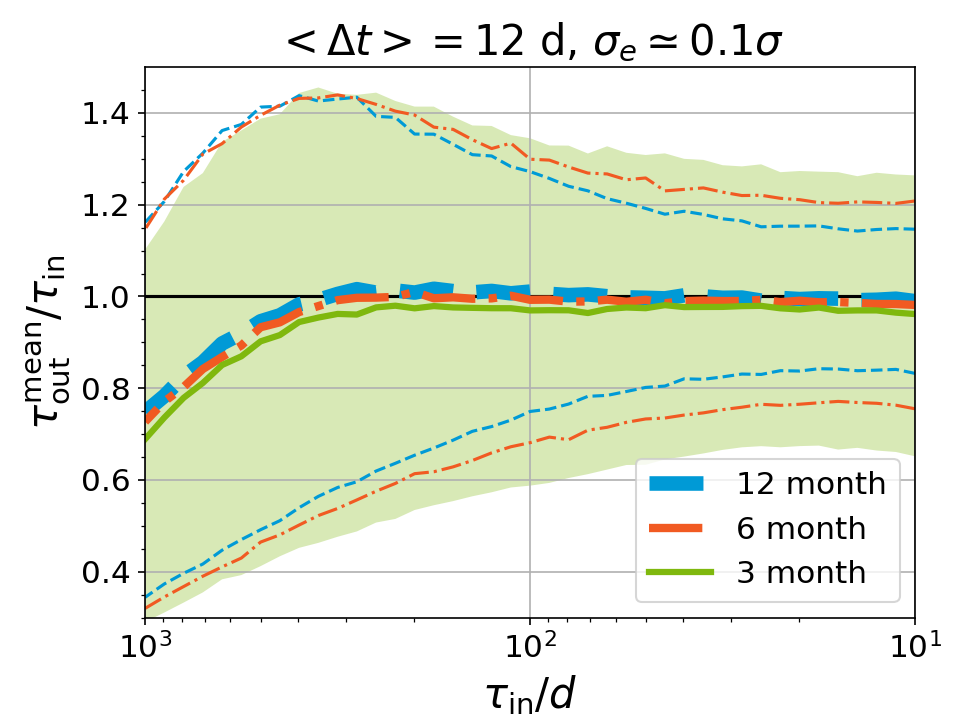}
     \caption{Similar to Figure~\ref{fig:epochs}, but for an AGN sample observed with $\sigma_{\rm e} \simeq 0.1\sigma$ and a fixed baseline of $T = 10$ year, unbiased timescales can be well retrieved for $\tau_{\rm in} \lesssim 0.1 T \simeq 300$ days, regardless of the quite sparse cadence $\langle \Delta t \rangle = 12$ days and of the limited observable duration of one year (i.e., 3 months and 6 months compared to the whole year). 
     }
     \label{fig:gap}
 \end{figure}

The minor effect of the average cadence on the retrieved timescale is clearly illustrated in Figure~\ref{fig:epochs}. For $T = 10 \tau_{\rm in}$, the timescales can be well retrieved as long as $\langle \Delta t \rangle < 0.2 \tau_{\rm in}$, with which the dispersion of retrieved timescales is independent of $\langle \Delta t \rangle$ (left panel of Figure~\ref{fig:epochs}). Instead, the dispersion can only be reduced by increasing the baseline, e.g., from $T = 10 \tau_{\rm in}$ to $100 \tau_{\rm in}$ (right panel of Figure~\ref{fig:epochs}). 

Moreover, when the baseline is $100 \tau_{\rm in}$, unbiased timescales can be obtained even with $\langle \Delta t \rangle \simeq 1 \tau_{\rm in}$. This is because when the baseline increases from $10 \tau_{\rm in}$ to $100 \tau_{\rm in}$ and the unbiased timescale is obtained at $\langle \Delta t \rangle$ increasing from $\simeq 0.2 \tau_{\rm in}$ to $\simeq 1\tau_{\rm in}$, the number of observed epochs indeed increases from 50 to 100 such that the short-term variations can still be somewhat covered by random observations.

Note that $\langle \Delta t \rangle \simeq 1 \tau_{\rm in}$ does not mean that there are no samplings with cadences smaller than $\tau_{\rm in}$. Instead, there are $\sim 65\%$ samplings with cadences smaller than $\tau_{\rm in}$. For comparison, there are $\sim 99\%$ samplings with cadences smaller than $\tau_{\rm in}$ for $\langle \Delta t \rangle \simeq 0.2 \tau_{\rm in}$.

\subsection{Season gap}

Real observations are affected by the season gap, which means that some AGNs can not be continuously observed throughout the year. To investigate the effect of season gap on the retrieved timescale, we simulate an AGN sample observed by 10 years, but within each year the observable duration is limited to 3 months and 6 months. A quite sparse cadence of $\langle \Delta t \rangle = 12$ days is selected such that AGNs are observed at $\simeq 8$, 15, and 30 epochs for 3, 6, and 12 months duration, respectively. For comparison, the well-known SDSS Stripe 82 is mapped 8 times on average in a 2-to-3 months duration per year and the average cadence is about 6 to 7 days \citep{MacLeod2010ApJ...721.1014M}, while a 3-day cadences is reached by the subsequent time domain surveys, such as Pan-STARRS1 (PS1; only in the median deep fields; \citealt{Chambers2016arXiv161205560C}) and ZTF \citep{Masci2019PASP..131a8003M}. The upcoming deep drilling fields of LSST will have a 2-day cadence \citep{Brandt2018arXiv181106542B}, and a 1-day cadence is proposed for the deep high-cadence survey of WFST \citep{WFST2023}.

As illustrated in Figure~\ref{fig:gap}, unbiased timescales can be well retrieved for $\tau_{\rm in} \lesssim 300$ days or again $\lesssim 0.1 T$, regardless of the presence of season gap. Indeed, larger dispersion is induced by the season gap. Owing to the given correlation between variance and timescale of AGN LC \citep{Kozlowski2021AcA....71..103K}, the retrieved timescales are slightly smaller than $\tau_{\rm in}$ as a result of the smaller variations implied by LCs with larger season gap. 

\begin{figure}[!t]
     \centering
     \includegraphics[width=0.45\textwidth]{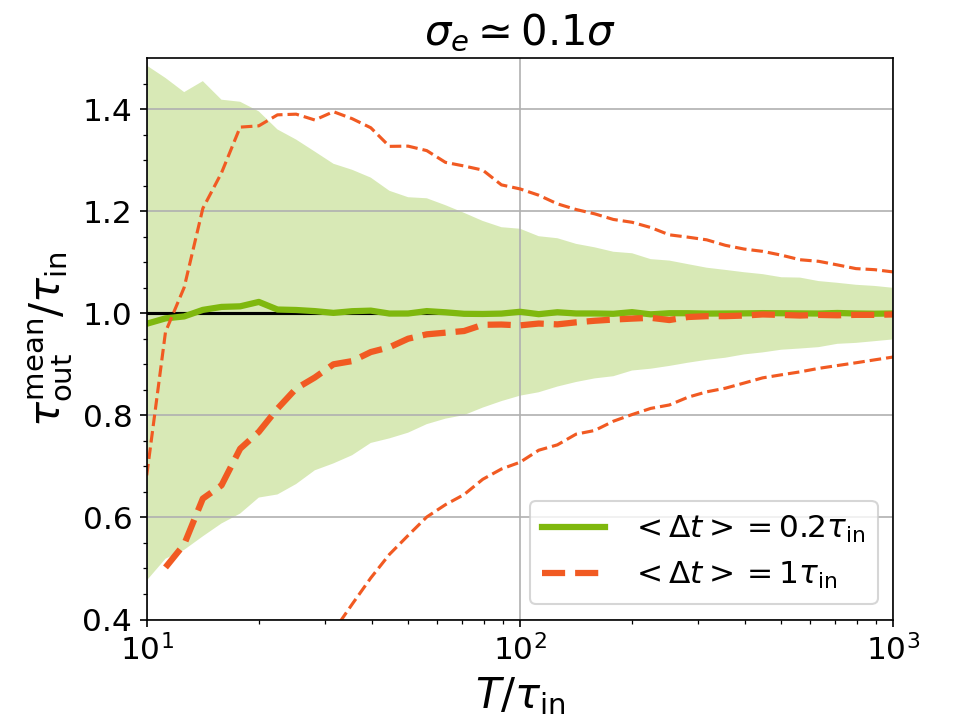}
     \caption{Similar to Figure~\ref{fig:epochs}, but for an AGN sample observed with $\sigma_{\rm e} \simeq 0.1\sigma$ and $\langle 
     \Delta t \rangle = 0.2 \tau_{\rm in}$ and $1 \tau_{\rm in}$, the retrieved timescale and the 16\%-84\% dispersion as a function of the baseline $T$ relative to $\tau_{\rm in}$.
     }
     \label{fig:cadence}
 \end{figure}

\subsection{Can the timescale of individual AGN be accurately determined?}

Unbiased timescales of an AGN sample can already be well retrieved given $T \simeq 10 \tau_{\rm in}$ and $\langle \Delta t \rangle \simeq 0.2 \tau_{\rm in}$, but with large dispersion (left panel of Figure~\ref{fig:epochs}). In other words, for individual AGN the retrieved timescale could be different from $\tau_{\rm in}$, e.g., by at most $\sim 50\%$ for a 68\% probability.

In terms of above analyses, the only effective way to reduce the dispersion of retrieved timescales is by increasing the baseline. In Figure~\ref{fig:cadence}, fixing $\sigma_{\rm e} \simeq 0.1\sigma$, dispersion of retrieved timescales is illustrated as a function of the baseline $T$ from $10 \tau_{\rm in}$ to $1000 \tau_{\rm in}$ for two cadences of $\langle \Delta t \rangle = 0.2 \tau_{\rm in}$ and $1 \tau_{\rm in}$. 
For $\langle \Delta t \rangle = 0.2 \tau_{\rm in}$, the dispersion is just gradually decreasing with increasing baseline. The 16\%-84\% dispersion of $\tau_{\rm out}/\tau_{\rm in}$ decreases from $\sim 50\%$ through $\sim 15\%$ to $\sim 5\%$ for baseline increasing from $10 \tau_{\rm in}$ through $100 \tau_{\rm in}$ to $1000 \tau_{\rm in}$, respectively. 

Even for $\langle \Delta t \rangle = 1 \tau_{\rm in}$, unbiased timescale can be obtained with $T \gtrsim 100 \tau_{\rm in}$ and the 16\%-84\% dispersion of $\tau_{\rm out}/\tau_{\rm in}$ can decrease from $\sim 20\%$ to $\sim 8\%$ for baseline increasing from $100 \tau_{\rm in}$ to $1000 \tau_{\rm in}$ (Figure~\ref{fig:cadence}). 
This has a practical implication that, for intermediate-mass BHs whose $\tau_{\rm in}$ are likely several days \citep{Burke2021Sci...373..789B}, continual monitoring up to several decades with $\langle \Delta t \rangle \sim \tau_{\rm in}$ would be sufficient for obtaining $\tau_{\rm in}$ with $\sim 20\%$ accuracy.

\subsection{The role of WFST in constraining the optical variation properties of AGNs}\label{sect:WFST}

\subsubsection{Survey Strategy of WFST}

Brief introductions to the WFST surveys and the relevant AGN sciences are presented in the following, while readers are referred to the WFST white paper for a panchromatic view \citep{WFST2023}. The 2.5-meter WFST with a field of view of 6.5 square degrees is designated to quickly survey the northern sky in four optical bands ($u$, $g$, $r$, and $i$). There will be two planned key programs across 6 years: a deep high-cadence $u$-band survey (DHS) and a wide field survey (WFS).

The DHS program tends to cover $\sim 720$ deg$^2$ surrounding the equator and reaches a depth of $\simeq 23$ and $\simeq 24$ mag (AB) in $u$ and $g$ bands in a 90-second exposure, respectively. Two separate DHS fields of $\sim 360$ deg$^2$ would be continuously monitored for 6 months per year in two to three bands (probably more $u$ and less $i$). At least 1-day cadence is allocated to each band (probably except $u$ around the full moon and $i$ around the new moon).
These quasi-simultaneous observations are dedicated to systematically unveiling the multi-band continuum lags of AGNs (Z. B. Su et al. 2023, in preparation).

The WFS program would cover $\sim 8000$ deg$^2$ in the northern sky and reach a depth of $\simeq 22.2$ and $\simeq 23.3$ mag (AB) in $u$ and $g$ bands in a 30-second exposure, respectively. Four separate WFS fields of $\sim 2000$ deg$^2$ would be continuously monitored for 3 months per year in the four bands. On average, a 6-day cadence is allocated to each band, or $\simeq 15$ visits per 3 months per band.

Both DHS and WFS are valuable for constraining the variability properties of AGNs, especially the $u$ band. According to the WFST schedule and the first light on September 17, 2023, we hereafter assume that the formal WFST scientific observation would start in spring 2024 and consider three WFST-extended baselines of 1, 6, and 10 years, coined as W1, W6, and W10, respectively.

\subsubsection{Beneficial complements to archive surveys}

The DHS footprint is planned to entirely cover the well-monitored SDSS Stripe 82 (S82) region ($\simeq 290$ deg$^2$; \citealt{MacLeod2010ApJ...721.1014M}), while the other DHS and WFS footprints would be enclosed by that of ZTF. The legacy surveys of WFST will provide time domain optical data with a cadence denser than SDSS S82 and a waveband shorter than ZTF.

In the following, we perform simulations by mocking WFST observations on AGNs to demonstrate how well can the extension of WFST baseline improve measuring variation properties of AGNs. 
We consider a total baseline of 24, 30, and 34 years, starting from the SDSS observations at around 2000 to 2024, 2030, and 2034 and including the 1, 6, and 10 years of WFST observations, respectively. On the basis of 9258 SDSS quasars in S82 region \citep{MacLeod2012ApJ...753..106M}, we search for their PS1 counterparts within one arcsec, resulting in 9254 quasars with 16 PS1 epochs complement to 62 SDSS epochs on average. 
Then we simulate 9254 LCs for any set of timescale ($\tau_{\rm in} = 10$ - 1000 days) and amplitude ($\sigma = 0.14$ mag), and implement the observed $r$-band epochs and photometric errors of 9254 quasars to the simulated 9254 LCs. 
Two relative photometric uncertainties ($\sigma_{\rm W} = 0.1 \sigma$ and $0.5 \sigma$) and two typical even cadences ($\Delta t_{\rm W} = 1$ day and 6 days for DHS and WFS, respectively) are considered for comparison.
Finally, variation properties of AGNs are retrieved using the afore-proposed optimized solution (i.e., K17PMm; Section~\ref{sec:style}). 
Note at the moment the similarity in $r$-band transmission curves among SDSS, PS1, and WFST leads to neglect the small corrections from SDSS or PS1 to WFST photometry.

\begin{figure*}[!t]
    \centering
    \includegraphics[width=0.45\textwidth]{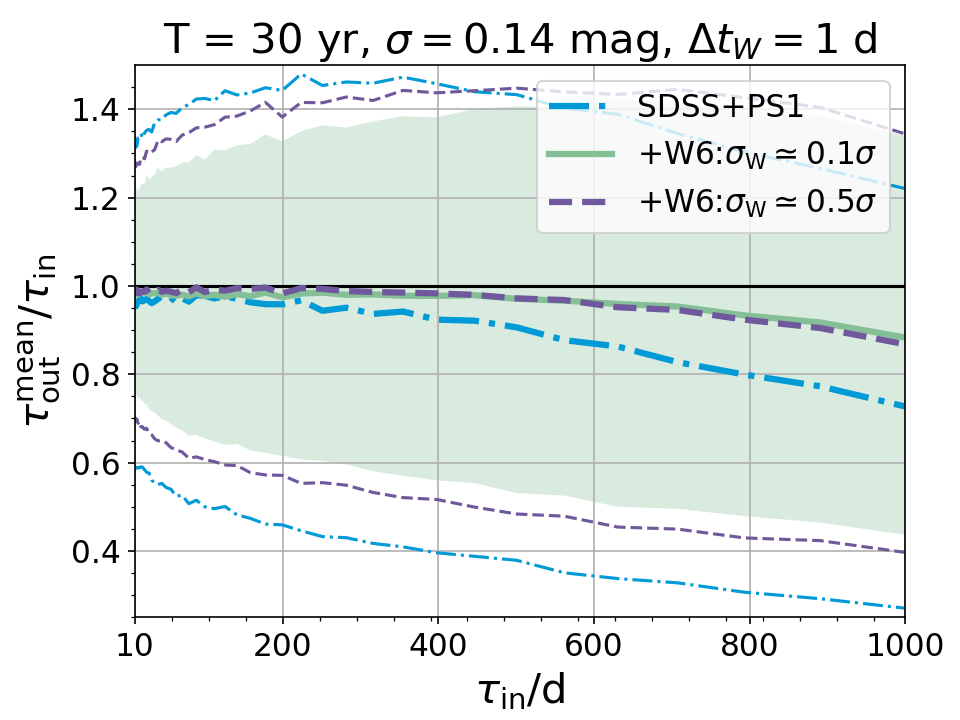}
    \includegraphics[width=0.45\textwidth]{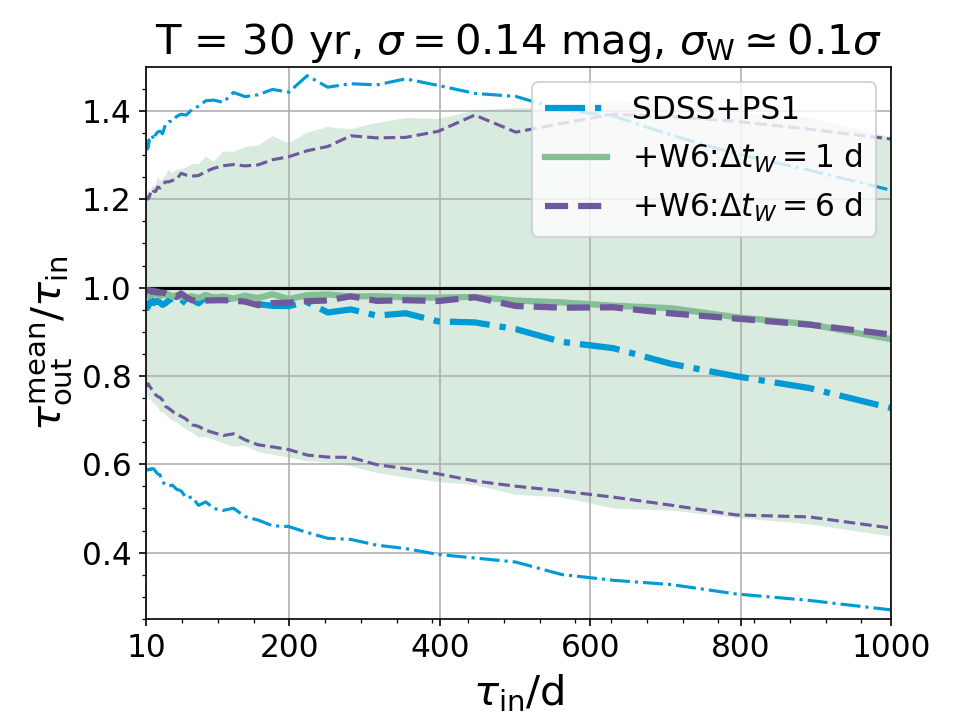}
    \includegraphics[width=0.45\textwidth]{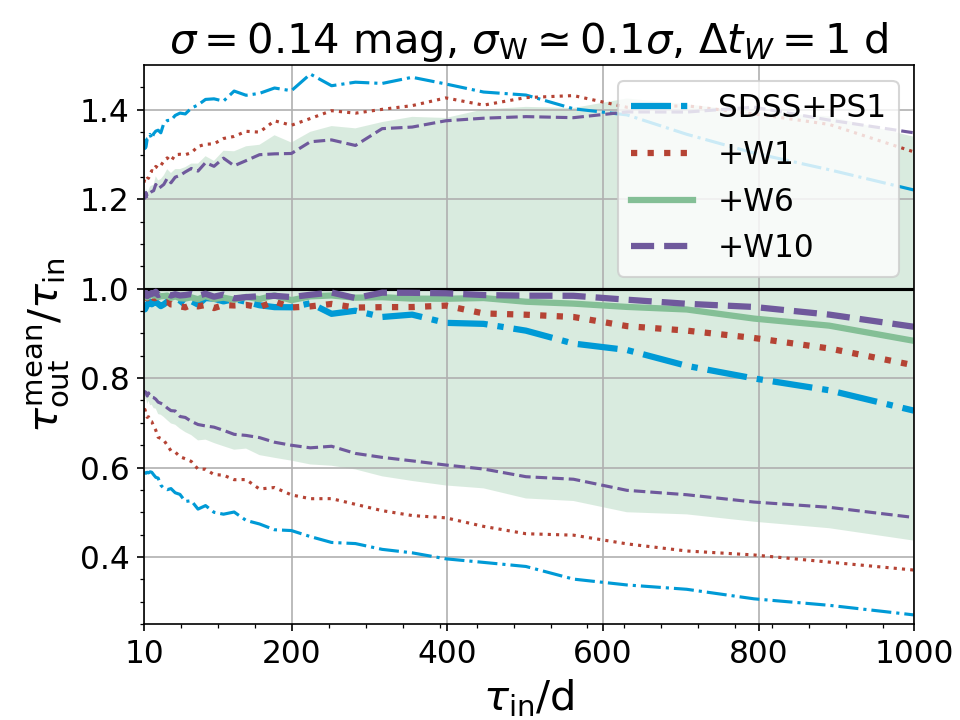}
    \caption{Adopting the real $r$-band cadences and photometry uncertainties of 9254 quasars in the S82 region with both SDSS and PS1 observations, we predict how well the retrieved timescales can be achieved by extending WFST observations, compared to solely considering SDSS+PS1 observations (dot-dashed lines). 
    For 9254 AGNs with $\sigma = 0.14$ mag and the fiducial WFST baseline of 6 years (W6; or a SDSS+PS1+W6 baseline of 30 years), we consider two WFST photometry uncertainties (i.e., $\sigma_{\rm W} \simeq 0.1 \sigma$ and $0.5 \sigma$; top-left panel) and two even cadences (i.e., $\Delta t_{\rm W} = 1$ day and 6 days; top-right panel). 
    Although dispersions (16\%-84\% percentile; green region or enclosed by two thin lines with the same style) of the retrieved timescales are larger for larger WFST photometry uncertainty, unbiased mean timescales can still be retrieved as long as $\simeq 400 - 600$ days for extending 6-year WFST observation (+W6). 
    Compared to +W6, extending the 1-year WFST observation (+W1) is already valuable, while the help of extending the 10-year WFST observation (+W10) is small, in retrieving unbiased mean timescales (bottom panel).
    }
    \label{4.2.1}
\end{figure*}

\begin{figure*}[!t]
    \centering
    \includegraphics[width=0.45\textwidth]{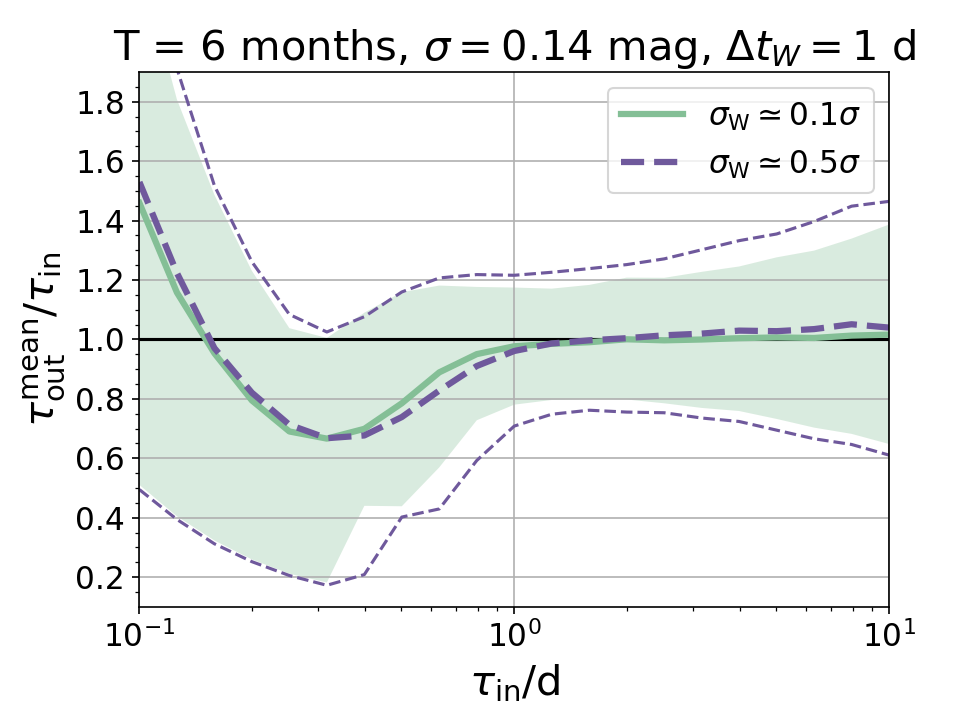}
    \includegraphics[width=0.45\textwidth]{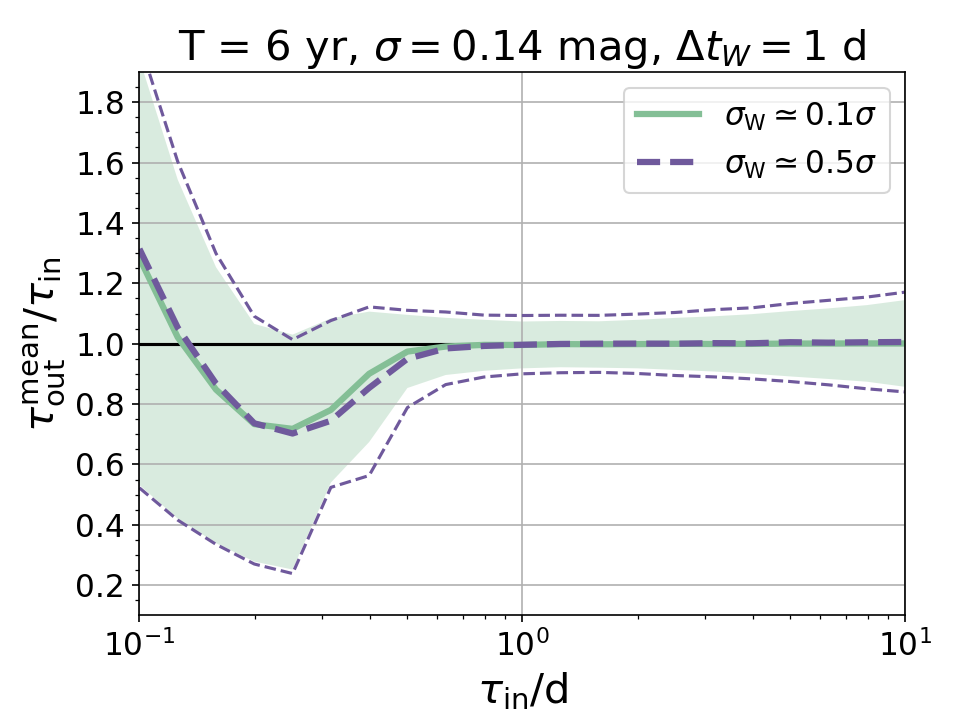}
    \caption{Left panel: for the WFST DHS with even cadence of $\Delta t_{\rm W} = 1$ day, unbiased timescales as short as $\simeq 1$ day can be retrieved, regardless of the photometry uncertainties, for the 1-year WFST DHS observation with a baseline of 6 months.
    Dispersions of retrieved timescales (16\%-84\% percentile) are only somewhat larger for larger relative photometry uncertainty.
    Right panel: similar to the left panel but for 6-month-per-year WFST DHS observations across 6 years. Accumulating WFST DHS observations from 1 year to 6 years, not only do the dispersions of the retrieved timescales get smaller but also the minimal unbiased timescale retrieved decreases to $\simeq 0.5$ day.
    }
    \label{4.3.1}
\end{figure*}

As illustrated in Figure~\ref{4.2.1}, unbiased mean timescales of quasars in the S82 region can be retrieved as long as $\simeq 400 - 600$ days for the fiducial 6-year WFST observation on the S82 region with a total baseline of $\sim 30$ years, regardless of the relative photometric uncertainties and cadences.
Compared to the 6-year WFST observation, the 1-year WFST observation is already valuable, as a result of nearly doubled baseline increasing from $\sim 13$ years (SDSS+PS1) to $\sim 24$ years (SDSS+PS1+W1), while the help of extending 6-year to 10-year WFST observations is small because the final baseline is further increased by only $\sim 13\%$.

\subsubsection{Retrieving unbiased timescales as short as about one day for intermediate-mass BHs in the WFST DHS}

Finding low-luminosity AGNs with intermediate-mass BHs (IMBHs) in dwarf galaxies is a challenge because the weak AGN signal is overwhelmed by the star-forming activity when conventional imaging and spectroscopic methods are adopted.
Instead, variability is probably a promising avenue for selecting IMBH candidates \citep{Baldassare2018ApJ...868..152B,Greene2020ARA&A..58..257G} and the measured timescale could be correlated with the BH mass of IMBH, though still subject to the lack of a sample of IMBHs with well measured BH masses in order to calibrate the intrinsic scatter of the $M_{\rm BH} - \tau$ relation at low luminosities \citep{Burke2021Sci...373..789B,Wang2023MNRAS.521...99W}. 

Nevertheless, to assess to what extent can the WFST DHS be used to retrieve unbiased timescales for IMBHs with $\sim 1 - 10$ days, probably corresponding to $M_{\rm BH} \sim 10^4 - 10^5\ M_\odot$ \citep{Burke2021Sci...373..789B}, we mock 6-month-per-year WFST DHS observations for IMBHs with $\tau_{\rm in} \sim 0.1 - 10$ days and fixed $\sigma = 0.14$ mag. Although some IMBHs can have variation amplitudes as large as $\sim 0.14$ mag, the typical variation amplitudes of IMBHs are likely $\sim 0.03$ mag or even smaller \citep{Baldassare2018ApJ...868..152B}. Therefore, we consider two WFST photometry uncertainties, $\sigma_{\rm W}$, relative to the fixed $\sigma$, that is, $\sigma_{\rm W} = 0.1 \sigma$ and $0.5 \sigma$.

The left panel of Figure~\ref{4.3.1} shows, with the 6-month WFST DHS observations in the first year survey, unbiased timescales of IMBHs from $\sim 1$ day to 10 days can be well retrieved, even for the case of $\sigma_{\rm W} \simeq 0.5\sigma$. 
If IMBHs within the WFST DHS fields are continually monitored for 6 years (the right panel of Figure~\ref{4.3.1}), dispersions of retrieved timescales will be smaller than $\sim 10\%$ and the minimal unbiased timescale decreases to $\simeq 0.5$ day. 
Therefore, we expect the WFST DHS would be helpful in retrieving accurate and unbiased timescales for individual IMBHs likely glowing as low-luminosity AGNs in dwarf galaxies.

\section{Summary}\label{sect:summary}

In this work, assuming the optical AGN variability is DRW-like, we examine how can the variation properties of AGNs be unbiasedly measured using the mainstream \textsc{celerite}.
We assess the effects of different priors and distinct ``best-fit'' values and find two best-combined estimators, that is, the K17 prior ($1/\sigma\sqrt{\tau}$) plus the ensemble mean of PM values (K17PMm solution) and the S21 prior ($1/\tau\sigma$) plus the ensemble median of PE values (S21PE50 solution). 
Since the MSE of the former is smaller, we propose the K17PMm solution as an optimized method to infer variation properties of AGNs and demonstrate that a minimal baseline 10 times longer than the input variation timescale is essential. 
Although the proposed solution may not perform well for the non-DRW-like AGN variability, our procedure in unveiling the corresponding optimized solution can be applied.

After scrutinizing our findings to those of \citet{Kozlowski2017}, \citet{Suberlak2021ApJ...907...96S}, and  \citet{Kozlowski2021AcA....71..103K}, we find different combinations of priors, ``best-fit'' values, and fitting methods indeed result in distinct conclusions on the minimal baselines required to retrieve unbiased variation timescales.
\citet{Kozlowski2017} use the PRH fitting method and also report a minimal baseline of $T/\tau_{\rm in} \sim 10$. We find the PRH fitting method should only be combined with another specific estimator, that is, the K17 prior plus the ensemble median of MAP values (K17MAP50 solution). Instead, our solution performs somewhat better than theirs and the utility of the fast \textsc{celerite} fitting method has a promising application in the time domain astronomy.

\citet{Suberlak2021ApJ...907...96S} suggest a shorter requested minimal baseline of $T/\tau_{\rm in} \sim 3 - 5$, but we can not confirm their finding by freeing the start points of simulated $10^4$ LCs. Therefore, we conclude the shorter minimal baseline suggested by them is attributed to the bias and randomness as a result of fixing the start point of simulated 100 LCs.

\citet{Kozlowski2021AcA....71..103K} claims a longer requested minimal baseline of $T/\tau_{\rm in} \sim 30$ is necessary and attributes the underestimation of timescales to the smaller variances of short LCs. Instead, our solution performs better on estimating the true long-term variability amplitude and so can shorten the requested minimal baseline to $T/\tau_{\rm in} \sim 10$.

Furthermore, utilizing the new optimized solution, we examine the impacts of several observational factors, including baseline $T$, number of epochs $N$ (cadence $\Delta t$), photometric uncertainty, and seasonal gap. Our findings are listed as follows:
\begin{enumerate}
    \item Larger photometric uncertainty only slightly increases the dispersion of retrieved timescales.
    \item The number of epochs should be sufficient, but excessive epochs are not very helpful. For $T = 10 \tau_{\rm in}$, timescales can be well retrieved as long as the average cadence $\langle \Delta t \rangle = 0.2 \tau_{\rm in}$. For a longer baseline $T = 100 \tau_{\rm in}$, a sparse average cadence $\langle \Delta t \rangle = \tau_{\rm in}$ is sufficient. Further decreasing cadences does not significantly reduce the dispersion of retrieved timescales.
    \item Seasonal gaps do not affect the result obviously, except for adding little dispersion and slightly depressing the ensemble mean of retrieved timescales as a result of smaller variance typically implied by LCs with larger season gaps.
    \item For sufficient small cadences, the retrieved timescale of individual AGN can reach an accuracy of $\sim 50\%$ ($1\sigma$ confidence level) for $T = 10 \tau_{\rm in}$, while extending baseline to $100 \tau_{\rm in}$ would increase the accuracy to $\sim 15\%$ ($1\sigma$ confidence level). 
\end{enumerate}

Finally, we apply our solution to evaluate constraints on the variation properties of AGNs to be provided by the WFST DHS. Complementing archive surveys in the SDSS S82 region, unbiased mean timescales of quasars covered by the WFST DHS can be retrieved as long as $\simeq 400 - 600$ days for the fiducial 6-year WFST DHS observation. Moreover, for the 1-year WFST observation, unbiased timescales as short as $\simeq 1$ day for IMBHs can be retrieved, while $\simeq 0.5$ day for the 6-year WFST DHS observation. Therefore, we expect the WFST DHS would be helpful in improving constraints on AGN variability.

\begin{acknowledgments}
We are grateful to the referee for valuable comments. We acknowledge S. Koz{\l}owski for his PRH code and a step-by-step instruction, and C. S. Kochanek, Yichen Jin, and Wenke Ren for their help.
This work is supported by the National Science Foundation of China (grant Nos. 12373016, 12033006, and 12192221), the USTC Research Funds of the Double First-Class Initiative with No. YD2030002009, and the Cyrus Chun Ying Tang Foundations.
\end{acknowledgments}

\appendix

\section{The PRH and CELERITE likelihood functions}\label{app:fit_methods}

Since developed by \citet[][PRH]{Press1992ApJ...385..404P} and upgraded by \citet{Rybicki1992ApJ...398..169R,Rybicki1995PhRvL..74.1060R}, the so-called PRH method, aiming at optimally reconstruct irregularly sampled LC, has been widely used to study AGN variability \citep{Kozlowski2010ApJ...708..927K,Kozlowski2016MNRAS.459.2787K,Kozlowski2017,Kozlowski2021AcA....71..103K} and incorporated into the well-known JAVELIN code by \citet{Zu2011ApJ...735...80Z,Zu2013ApJ...765..106Z,Zu2016ApJ...819..122Z}. 
Briefly, if the measured LC $\itbf{y}(t)$ with $N$ data points is composed of a true signal $\itbf{s}(t)$, a measurement noise $\itbf{n}(t)$, and a response matrix $\itbf{L}$ defining a general trend with a set of linear coefficients $\itbf{q}$, we have $\itbf{y} = \itbf{s} + \itbf{n} + \itbf{L} \itbf{q}$.
Here $\itbf{y}$, $\itbf{s}$, and $\itbf{n}$ are all $N \times 1$ matrices. If there are $M$ coefficients in $\itbf{q}$, $\itbf{q}$ is thus a $M \times 1$ matrix and $\itbf{L}$ is a $N \times M$ matrix.
For our purpose, $\itbf{q}$ has only one element, $m$, indicating the mean of the LC and thus $\itbf{L}$ is a $N \times 1$ matrix with all elements equal to one.
The PRH likelihood to be maximized to measure the model parameters (e.g., $\tau$ and $\sigma$) is
\begin{equation}
    {\cal L}_{\rm PRH}(\itbf{y} | \itbf{s}, \itbf{n}, \tau, \sigma) \propto |\itbf{C}|^{-1/2} |\itbf{L}^{T} \itbf{C}^{-1} \itbf{L}|^{-1/2} \exp\left(- \frac{\itbf{y}^{T} \itbf{C}^{-1}_{\bot} \itbf{y}}{2} \right),
\end{equation}
where the total covariance matrix is $\itbf{C} = \itbf{S} + \itbf{N}$, $\itbf{S}(\itbf{s}, \tau, \sigma)$ is the covariance matrix of $\itbf{s}$ given model parameters, $\itbf{N}$ is the covariance matrix of $\itbf{n}$, and $\itbf{C}^{-1}_{\bot} = \itbf{C}^{-1} - \itbf{C}^{-1} \itbf{L} (\itbf{L}^{T} \itbf{C}^{-1} \itbf{L})^{-1} \itbf{L}^{T} \itbf{C}^{-1}$. Further assuming a specific covariance function, such as the one for the DRW process (Equation~\ref{eq:drw_cov}), the $ij$-element $S_{ij}$ of $\itbf{S}$ is determined by ${\rm Cov}(\Delta t_{ij}, \tau, \sigma) = \sigma^2 \exp(- \Delta t_{ij} / \tau)$, where $\Delta t_{ij} = |t_i - t_j|$ and $t_i$ (or $t_j$) is the time at the $i$-epoch (or $j$-epoch). For uncorrelated noise, the non-zero elements of $\itbf{N}$ are $N_{ii} = n_i^2$, where $n_i$ is the $i$-th element of $\itbf{n}$.

Instead, the likelihood implemented in \textsc{celerite} \citep{Foreman-Mackey2017AJ....154..220F,Aigrain2023ARA&A..61..329A} is 
\begin{equation}
    {\cal L}_{\rm CELERITE}(\itbf{y} | \itbf{s}, \itbf{n}, \tau, \sigma, m) \propto |\itbf{C}|^{-1/2} \exp\left( - \frac{ \itbf{r}^{T} \itbf{C}^{-1} \itbf{r} }{2} \right),
\end{equation}
where $\itbf{r} = \itbf{y} - m$. Here, freeing $m$ or fixing it to the mean of the simulated LC does not have significant effects on the other two parameters.

\begin{figure}[!t]
    \centering
    \includegraphics[width=0.6\textwidth]{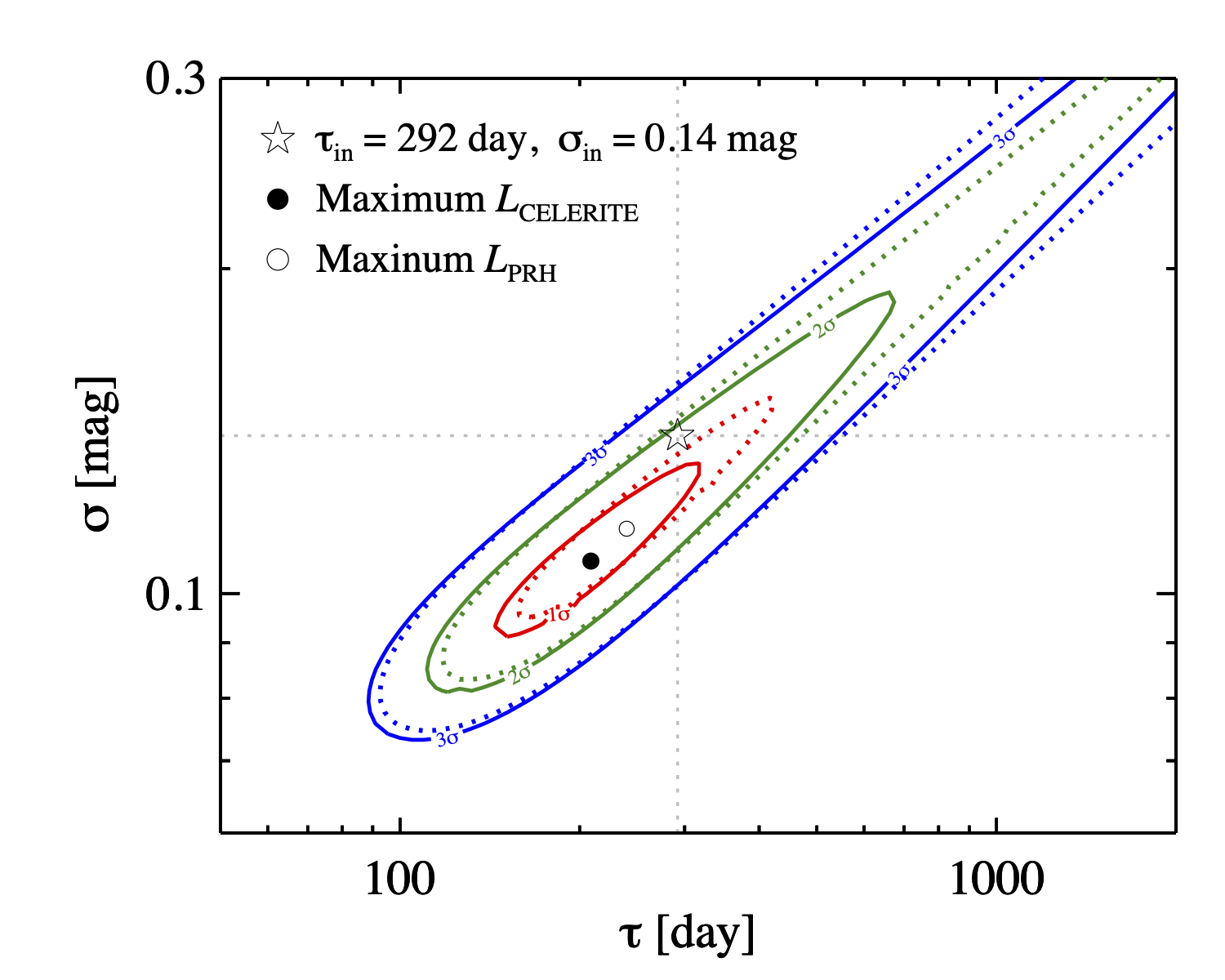}
    \caption{Given a typical simulated LC of an OGLE-observed AGN with $\tau_{\rm in} = T / 10 = 292$ day for $T = 8$ years and $\sigma_{\rm in} = 0.14$ mag (open star for the input parameters), the likelihood surfaces (i.e., red, green, and blue contours for $\Delta \ln {\cal L} = -0.5$, -2.0, and -4.5 or, equivalently, 1, 2, and $3\sigma$ confidence levels) implied by the \textsc{celerite} (solid contours) and PRH (dotted contours) likelihood functions are compared together with their maximum likelihood points, i.e., filled and open circles, respectively.}
    \label{fig:posterior_comparison}
\end{figure}

To illustrate the difference between these two likelihoods, we show in Figure~\ref{fig:posterior_comparison} the two-dimensional likelihood surfaces implied by them for a given typical simulated LC of an OGLE-observed AGN with $\tau_{\rm in} = T/10$. Compared to the \textsc{celerite} likelihood, the PRH likelihood prefers larger $\tau$ and $\sigma$ with a clear longer tail toward the up-right corner, primarily because the PRH likelihood function contains a unique term, $|\itbf{L}^{T} \itbf{C}^{-1} \itbf{L}|^{-1/2} \propto |\itbf{C}| \propto \sigma^2 \exp(-1/\tau)$, which increases the likelihood for larger $\tau$ and $\sigma$. This thus accounts for the difference of parameters retrieved adopting the K17MAP50 solution shown in the left panel of Figure~\ref{fig:K17_S21_comparison_tau}.

\section{Additional white noise}\label{app:extra_err}

\begin{figure}[!ht]
    \centering
    \includegraphics[width=0.3\textwidth]{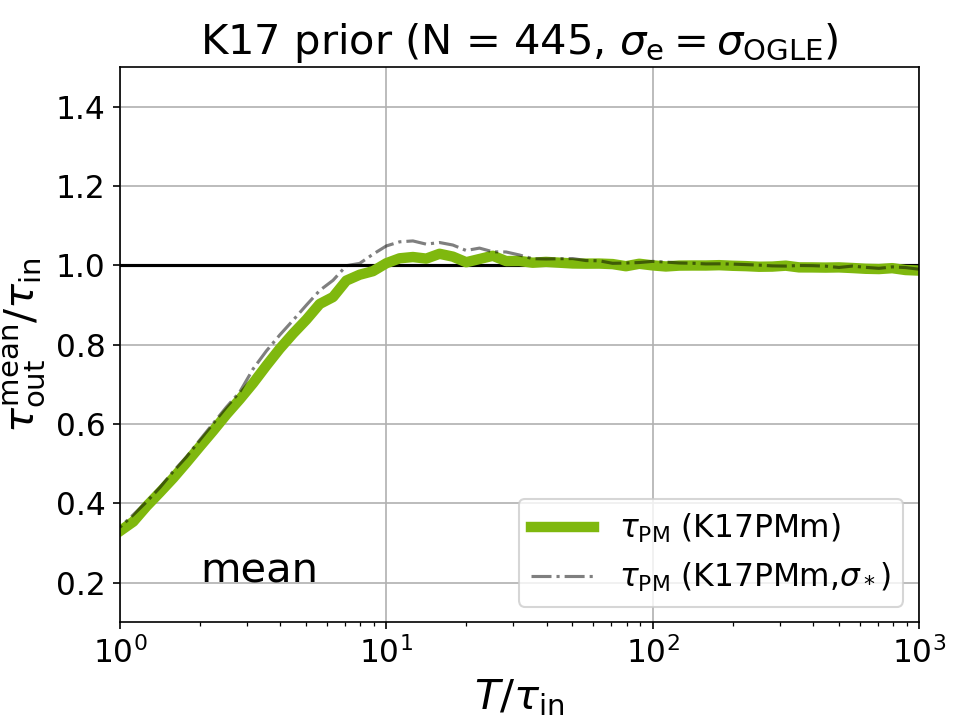}
    \includegraphics[width=0.3\textwidth]{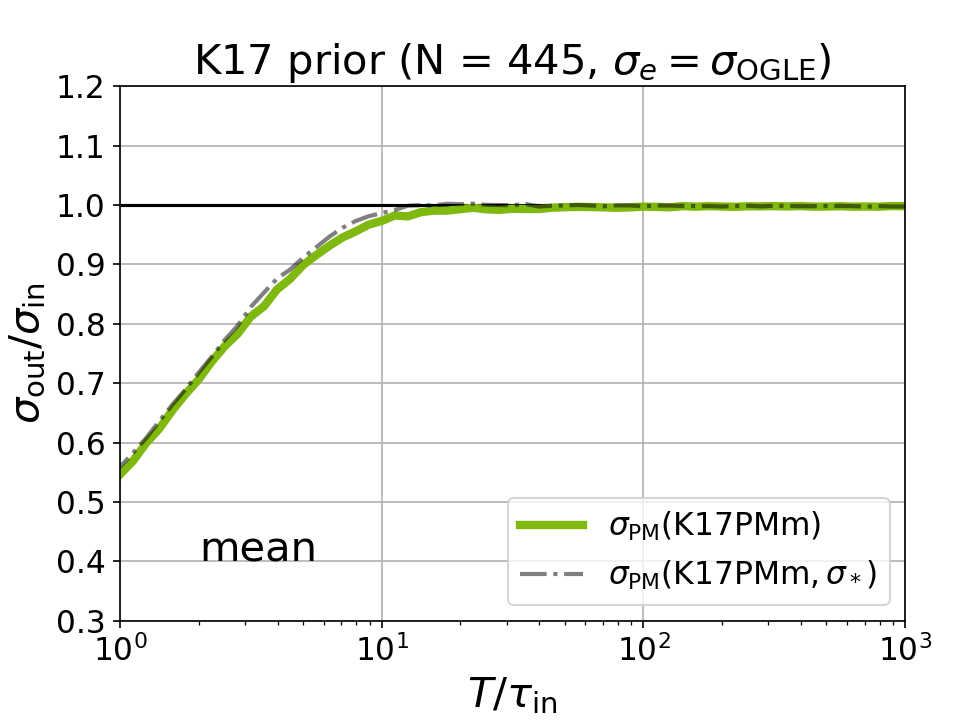}
    \includegraphics[width=0.3\textwidth]{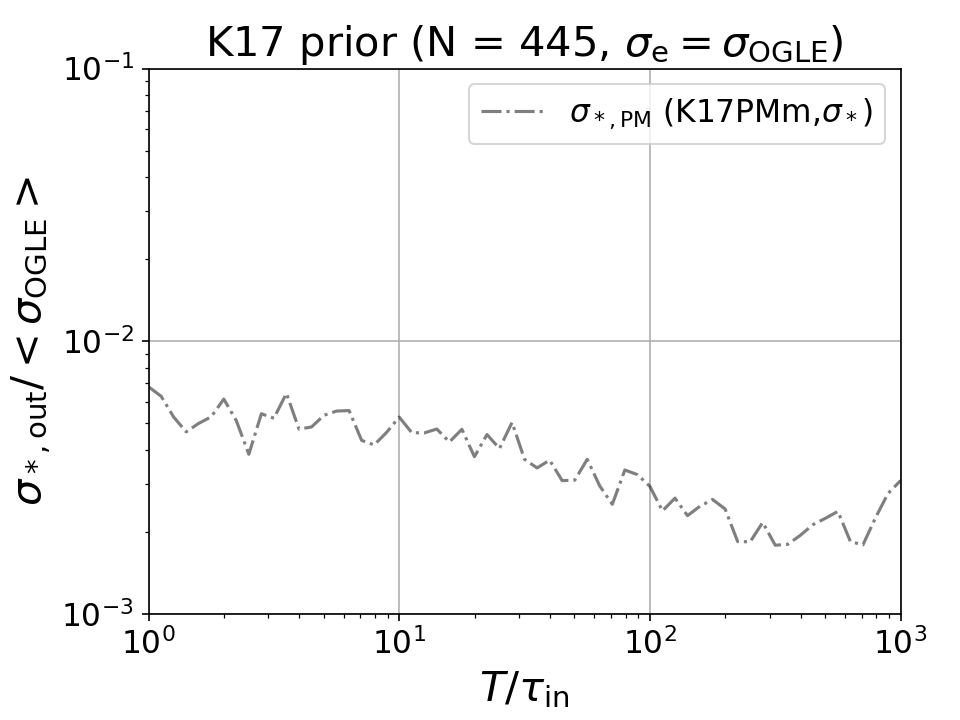}
    \caption{Left panel: same as the top-left panel of Figure~\ref{fig:tau_out_in_n445}, but only the variation timescale retrieved by the K17PMm solution (green solid line) is shown and compared to that retrieved including $\sigma_*$ (grey dot-dash line). 
    Middle panel: same as the left panel but for the retrieved variation amplitude $\sigma$. 
    Right panel: adopting the same K17PMm solution, illustrated is the retrieved $\sigma_*$ relative to the average OGLE photometric uncertainty (i.e., $\langle \sigma_{\rm OGLE} \rangle \simeq 0.025$ mag) as a function of $T/\tau_{\rm in}$.
    }
    \label{fig:extra_err}
\end{figure}

An additional white noise term, $\sigma^2_*$, such that $N_{ii} = n_i^2 + \sigma^2_*$, could be introduced to account for additional sources of photometric errors that have not been included in the measured errors \citep[e.g.,][]{Burke2021Sci...373..789B,Wang2023MNRAS.521...99W}.
It should be noted that this term assumes a constant unknown error for every epoch and it is only able to deal with the case where photometric errors are underestimated. To examine whether adding this term would affect our result, we take the K17PMm solution and find small differences, $\lesssim 4\%$, between model parameters retrieved without and with $\sigma_*$ (Figure~\ref{fig:extra_err}). 
The best-fit $\sigma_*$ is also very small with a ratio of $\sigma_*$ to the average $\sigma_{\rm OGLE}$ only several times $10^{-3}$. Therefore, we conclude that adding an extra noise term is unnecessary for us and has little effect on our conclusion.

\bibliographystyle{aasjournal}
\bibliography{ms.bbl}

\end{document}